\documentclass[useAMS,usenatbib]{mn2e}
\usepackage{amssymb}
\usepackage{graphicx}

\usepackage{ifthen}
\def\draftversion{1} 

\ifthenelse{\equal{\draftversion}{1}}{
	\usepackage{xcolor}
	\newcommand{\tmp}{}
	\newenvironment{envcomm}[1]{\renewcommand{\tmp}{#1}\begin{color}{blue}\begin{center}\hrule\vspace{0.5mm}\tmp's COMMENTS\end{center}}{\begin{center}END OF \tmp's COMMENTS\vspace{0.5mm}\hrule\end{center}\end{color}}
	\newenvironment{draft}{\begin{color}[rgb]{0,0.4,0}\begin{center}\hrule\vspace{0.5mm}DRAFT\end{center}}{\begin{center}END OF DRAFT\vspace{0.5mm}\hrule\end{center}\end{color}}
	\newcommand{\comcomm}[2]{\begin{color}{blue}\ $\bullet$ \textbf{#1:} #2 $\bullet$\ \end{color}}
	\newcommand{\revend}[1]{\par\begin{color}[rgb]{0,0.4,0}\begin{center}\hrule\vspace{0.5mm}END OF #1's REVISIONS\vspace{0.5mm}\hrule\end{center}\end{color}\par}
	\newcommand{\todo}[1]{\begin{color}{red}\ $\bullet$ \textbf{To do: }#1 $\bullet$\ \end{color}}
	
	\newcommand{\del}[1]{\begin{color}[rgb]{0,0.5,0.0}\ $\bullet$ \textbf{Deleted: }#1 $\bullet$\ \end{color}}
	\newcommand{\sk}[1]{\begin{color}[rgb]{0.6,0,0.6}#1\end{color}}
	}{
	\newsavebox{\trashcan}
	\newenvironment{envcomm}[1]{\begin{lrbox}{\trashcan}\begin{minipage}{\columnwidth}}{\end{minipage}\end{lrbox}}
	
	\newcommand{\comcomm}[2]{}
	\newcommand{\revend}[1]{}
	\newcommand{\todo}[1]{}
	
	\newcommand{\del}[1]{}
	\newcommand{\sk}[1]{}
	}

\newcommand{\araa}{ARA\&A}
\newcommand{\apj}{ApJ}
\newcommand{\apjl}{ApJ}
\newcommand{\apjs}{ApJS}
\newcommand{\aap}{A\&A}
\newcommand{\aapr}{A\&A~Rev.}
\newcommand{\mnras}{MNRAS}

\newcommand{\mh}{\ensuremath{\textrm{\,--\,}}}
\newcommand{\bb}[1]{\ifmmode \mbox{\boldmath $ #1$} \else  \mbox{\boldmath $#1$} \fi}

\newcommand{\dd}{\ensuremath{\,\mathrm{d}}}

\newcommand{\U}[1]{\ensuremath{\mathrm{~#1}}}
\newcommand{\erg}{\U{erg}}

\newcommand{\yr}{\U{yr}}
\newcommand{\Myr}{\U{Myr}}
\newcommand{\Gyr}{\U{Gyr}}
\newcommand{\pc}{\U{pc}}
\newcommand{\kpc}{\U{kpc}}

\newcommand{\msun}{\U{M}_{\odot}}
\newcommand{\Msun}{\msun}
\newcommand{\cc}{\U{cm^{-3}}}
\newcommand{\kms}{\U{km\ s^{-1}}}

\newcommand{\hi}{H{\sc i} }
\newcommand{\hii}{H{\sc ii} }
\newcommand{\mach}{\ensuremath{\mathcal{M}}}
\newcommand{\tff}{\ensuremath{t_\mathrm{ff}}}

\newcommand{\cellsize}{\ensuremath{d_x}}
\newcommand{\cellmin}{\ensuremath{d_{x,\mathrm{min}}}}

\newcommand{\vrot}{\ensuremath{v_\mathrm{rot}}}
\newcommand{\vcirc}{\ensuremath{v_\mathrm{circ}}}
\newcommand{\vdisp}{\ensuremath{\sigma_\mathrm{v}}}

\newcommand{\mhii}{\ensuremath{M_\mathrm{HII}}}
\newcommand{\rhii}{\ensuremath{r_\mathrm{HII}}}
\newcommand{\thii}{\ensuremath{T_\mathrm{HII}}}
\newcommand{\etaob}{\ensuremath{\eta_\mathrm{OB}}}
\newcommand{\omegap}{\ensuremath{\Omega_\mathrm{p}}}
\newcommand{\rr}{\ensuremath{\alpha_\mathrm{r}}}
\newcommand{\sigmapdf}{\ensuremath{\sigma_\mathrm{PDF}}}
\newcommand{\multiscat}{\ensuremath{s}}

\newcommand{\fig}[2][]{Figure#1~\ref{fig:#2}}

\renewcommand{\fig}[2][]{Fig#1.~\ref{fig:#2}}

\title[A sub-parsec resolution simulation of the Milky Way]{A sub-parsec resolution simulation of the Milky Way: Global structure of the ISM and properties of molecular clouds}

\author[Renaud et al.]{F.~Renaud$^1$\thanks{florent.renaud@cea.fr}, F.~Bournaud$^1$, E.~Emsellem$^{2,3}$, B.~Elmegreen$^4$, R.~Teyssier$^{1,5}$,
\newauthor J.~Alves$^6$, D.~Chapon$^1$, F.~Combes$^7$, A.~Dekel$^8$, J.~Gabor$^1$,
\newauthor P.~Hennebelle$^{1}$, K.~Kraljic$^1$\\
\\
$^1$ Laboratoire AIM Paris-Saclay, CEA/IRFU/SAp, Universit\'e Paris Diderot, F-91191 Gif-sur-Yvette Cedex, France\\
$^2$ European Southern Observatory, 85748 Garching bei Muenchen, Germany\\
$^3$ Universit\'e Lyon 1, Observatoire de Lyon, CRAL et ENS, 9 Av Charles Andr\'e, F-69230 Saint-Genis Laval, France\\
$^4$ IBM T. J. Watson Research Center, 1101 Kitchawan Road, Yorktown Heights, New York 10598 USA\\
$^5$ Institute for Theoretical Physics, University of Z\"urich, CH-8057 Z\"urich, Switzerland\\
$^6$ Institute for Astronomy, University of Vienna, T\"urkenschanzstrasse 17, 1180 Vienna, Austria\\
$^7$ Observatoire de Paris, LERMA et CNRS, 61 Av de l'Observatoire, F-75014 Paris, France\\
$^8$ Racah Institute of Physics, The Hebrew University, Jerusalem 91904, Israel
}

\date{Accepted 2013 September 6.  Received 2013 September 3; in original form 2013 May 28}

\begin{document}
\maketitle

\begin{abstract}
We present a self-consistent hydrodynamical simulation of a Milky Way-like galaxy, at the resolution of $0.05 \pc$. The model includes star formation and a new implementation of stellar feedback through photo-ionization, radiative pressure and supernovae. The simulation resolves the structure of the interstellar medium at subparsec resolution for a few cloud lifetimes, and at $0.05 \pc$ for about a cloud crossing time. Turbulence cascade and gravitation from the kpc scales are \emph{de facto} included in smaller structures like molecular clouds. We show that the formation of a bar influences the dynamics of the central $\sim 100 \pc$ by creating resonances. At larger radii, the spiral arms host the formation of regularly spaced clouds: beads on a string and spurs. These instabilities pump turbulent energy into the gas, generally in the supersonic regime. Because of asymmetric drift, the supernovae explode outside of their gaseous nursery, which diminishes the effect of feedback on the structure of clouds. The evolution of clouds is thus mostly due to fragmentation and gas consumption, regulated mainly by supersonic turbulence. The transition from turbulence supported to self-gravitating gas is detected in the gas density probability distribution function at $\sim 2000 \cc$. The power spectrum density suggests that gravitation governs the hierarchical organisation of structures from the galactic scale down to a few parsecs. 
\end{abstract}

\section{Introduction}

Stars form out of the gaseous medium of galaxies. This well-known fact hides a collection of complex and not yet perfectly understood details, owing to the diversity of physical conditions one may find in galaxies, and even within a given galaxy. The gaseous structures are being probed deeper and deeper, providing a mine of information on the fine details of star formation. For instance, recent Herschel observations of the Milky Way interstellar medium (ISM) revealed a filamentary structure of the star forming regions at sub-parsec scale \citep{Arzoumanian2011}. These filaments might be formed by turbulent shocks, during the fragmentation of their host molecular clouds \citep{Padoan2001}. But like star formation, turbulence remains a poorly understood topic.

The turbulence in galaxies seems to be generated by large scale ($> 1 \kpc$) motions of the ISM and its energy is transferred down to the small scales by the so-called turbulence cascade \citep[e.g.][]{Padoan2009, Bournaud2010b}, where it is dissipated into heat. Because of this scale coupling, understanding the process of star formation requires a description of the ISM at sub-parsec resolution \emph{but} in its galactic context. On the one hand, cosmological aspects like accretion of gas \citep[e.g.][]{Dekel2006} or galaxy interaction/mergers \citep{diMatteo2007, Karl2010, Teyssier2010} are known to affect the behaviour of the gas reservoir of the galaxy, and thus the formation of stars. On the other hand, newly born stars themselves modify the properties of the gas in their vicinity through feedback. Numerically, a fully consistent description of these effects spanning at least $8\mh 9$ orders of magnitude in space is still out-of-reach: because of technical limitations, simulations probing star formation face either a problem of resolution or a lack of physical ingredients implemented. 

To date, the problem has been tackled from both ends: large scale ($\gtrsim 100 \kpc$) simulations covering a long time lapse ($\gtrsim 1 \Gyr$) concentrate on galaxy formation and evolution, while small scale ($\lesssim 1 \pc$) relatively short ($\lesssim 1 \Myr$) runs focus on stellar populations. For example, the connection to cosmology has been addressed by selecting halos at high redshift in cosmological simulations, and focussing on the assembly of one galaxy over a very long time, but at the cost of a spatial resolution limited to a few $100 \pc$ \citep[see][for a simulation of a Milky Way-like galaxy]{Guedes2011}. At slightly smaller scale, a number of studies have shown the effect of galaxy interactions and inner structures (spirals, bars) in the evolution of the star formation rate \citep{diMatteo2007,Dobbs2008}. Such works model entire galaxies and thus properly describe the environment of star forming regions, but use sub-grid recipes to implement star formation and form stellar \emph{particles} representing more than one star, from a few $10^3$ to $10^6 \Msun$, the structure of the ISM at individual star formation scale being unresolved. In particular, \citet{Tasker2011} explored the connection between the formation of giant molecular clouds (GMCs) and their surrounding medium, underlining the role of cloud-cloud mergers and star formation in the early destruction of clouds. Similarly, \citet{Dobbs2012} emphasized the role of stellar feedback on the formation/destruction cycle of GMCs in the context of spiral galaxies, and \citet{Hopkins2012} focussed on a multi-component stellar feedback in several galaxy types, including models of high redshift discs ($z\sim 2$).

On the stellar side, the formation of individual stars is followed from the fragmentation of an isolated molecular cloud \citep{Bate2005, Federrath2012}. Although the collisional aspect of these objects is often neglected at the price of missing the evolution of binary/multiple stars, the clustered formation of a few of tens $\sim 1 \msun$ particles is well described. However, in this case the environmental effect of the galaxy is not present or is highly idealised. Studies where a time-dependent potential mimicking the gravitational well of a spiral arm has been included have been performed \citep[e.g.][]{Bonnell2006}, but the creation of the cloud itself and the fully consistent injection of turbulence from the galactic context were still missing.

\citet{Bonnell2013} started to bridge the gap by considering the influence of kpc-scale dynamics on star formation using to a two-step method. First, a ``low'' resolution simulation of spiral arms is performed. Then a re-simulation of one of the dense clouds formed is done at much higher resolution. This innovative method allowed to approach the resolution of individual star formation thanks to sink particles \citep{Bate1995} accreting several tens of gas particles ($\sim 0.1 \Msun$) and creating $\approx 11 \Msun$ stars. Despite the giant leap made in the coupling of galactic and stellar scales, this work lacks the fully consistent galactic dynamics during the re-simulations and the influence of stellar feedback, that might be of significant importance at both scales.

In the present work, we take one more step toward the goal of merging the two approaches by increasing the resolution of a galaxy simulation down to subparsec scales. Yet limited by currently available computer ressources, we could not probe processes at pre-stellar core scale: our results still rely on sub-grid recipes. However, the resolutions reached in space, mass and temperature allow us to properly describe the structure of the interstellar medium at the scales of star forming cores ($\sim 0.05 \pc$) in the \emph{fully self-consistent} context of an isolated galaxy ($\sim 100 \kpc$). The (dynamic) gravitational potential, gas flows and the turbulence are \emph{de facto} included at the scale of star formation. Our model reproduces the structures of a grand-design spiral galaxy tailored to ressemble the Milky Way, via the Besan\c{c}on model \citep{Robin2003}, although a one-to-one match is not sought.

This paper first presents the numerical method used in Section~\ref{sec:method}. Section~\ref{sec:ic} details the initial conditions and the preliminary evolution of the model leading to a grand-design spiral galaxy with many substructures. The properties of the large-scale gaseous structures are explored in Section~\ref{sec:largescale}. Star formation, gas flows in the bar and internal structures of star forming clumps are topics deserving a contribution of their own, and will be presented in several forthcoming papers.

\section{Numerical method}
\label{sec:method}

\subsection{Code, methods and ressources}
\label{sec:ramses}

The simulation is performed with the Adaptive Mesh Refinement (AMR) code RAMSES \citep{Teyssier2002}. For simplicity, we chose to ignore the magnetic fields. While some studies speculated they might play an important role in supporting the gas against collapse \citep{Shu1987,VanLoo2013}, recent works showed that turbulence is the most important factor, while magnetic fields have a more moderate but still significant influence on the star formation efficiency and the dense gas mass fraction \citep{Padoan2011,Federrath2012,Federrath2013,Kainulainen2013}. We also note that including magnetic fields could modify the morphology of spiral arms \citep{Pakmor2013} and their collapse into clouds via magneto-Jeans instabilities \citep{Kim2002}.

The dark matter halo, the old primordial stars, the new stars born from the conversion of gas (see Section~\ref{sec:sf}) and the central supermassive black hole are evolved using a particle-mesh solver.

On top of the particles, the evolution of the gaseous component is computed by solving the Euler equations on the adaptive grid with a second-order Godunov scheme, and an acoustic Riemann solver with a ``minmod'' slope limiter\footnote{by setting RIEMANN=acoustic and SLOPE\_TYPE=1 (see the RAMSES user guide for details).}. The AMR technique is used to put the highest resolution on the densest regions of the galaxy, while keeping a low, computationally cheap, resolution in the low density volumes. The refinement strategy is based on the density of the baryons (stars and gas) and on their stability: a thermal Jeans length is always resolved by at least four cells (see Section~\ref{sec:eos}). The simulation box spans $100 \kpc \times 100 \kpc \times 100 \kpc$ with a coarsest mesh covering the entire volume made of $200 \pc$ wide cells, and the finest cells of $0.05 \pc$ in the most refined volumes. Note however that any study on the structure or the evolution of objects requires to resolve them with a few elements (cells, particles, pixels). Therefore, only the exploration of objects spanning a few $\times 0.05 \pc$ are within reach of this simulation. The finest cells are created on the 21st level of refinement. Although the exact number of cell varies with the adaptive refinement, depending on the structure of the galaxy, the simulation counts about 240 millions of them, on average.

Contrary to the new formed stars which are created in compact, dense clouds, the gravitational field of the pre-existing stars and the dark matter contains less small scale features and does not require description at the same accuracy. To avoid the numerical cost of a large number of particles, this field is rendered by only 60 million particles and is smoothed over $3 \pc$. However, gravitation of the new stars and self-gravity of the gas component is evaluated at the full resolution of the grid, i.e. down to $0.05 \pc$.

The simulation has been run for 12 million core-hours on 6080 cores on the ``fat nodes'' of the supercomputer \emph{Curie} hosted at the \emph{Tr\`es Grand Centre de Calcul} (TGCC). Because of the computational cost of such a simulation, variations of the parameters and methods and a statistical approach are yet not possible, and only one run has been performed. However, a wide diversity of gas clouds allows a comparison of their formation, structure and evolution under various physical conditions.

\subsection{Equation of state}
\label{sec:eos}

\begin{figure}
\includegraphics{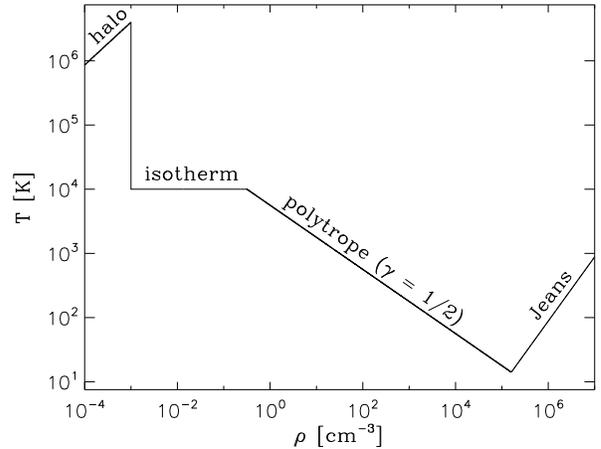}
\caption{Effective equation of state, including the Jeans polytrope to avoid artificial fragmentation in the smallest cells.}
\label{fig:eos}
\end{figure}

In simulations like this one, a very small fraction of the gas can temporarily become very hot, usually well above/below the plane of the disc. Since such a hot gas requires to set a small timestep to fulfill the \citet*{Courant1928} condition, this represents a significant slowing down of the simulation. To avoid this, we replace the evaluation of the heating and cooling processes due to ultraviolet background and atomic/molecular lines \citep{Haardt1996}, with a piecewise polytropic equation of state (EOS) derived from a fit of actual heating/cooling balances at 1/3 solar metalicity \citep{Bournaud2010b}, and plotted in \fig{eos}. (Comparisons between the use of this approximation and a more precise recipe will be presented in a forthcoming contribution, Kraljic et al., in prep.)

The EOS is isothermal ($10^4 \U{K}$) for the gas with densities $10^{-3} \cc < \rho < 0.3 \cc$, and is a polytrope $T\propto \rho^{-1/2}$ for the denser regions. Because the resolution is finite, the subgrid physics is accounted for by artificially adding a pressure floor in the densest cells, according to \citet[see also \citealt{Robertson2008}]{Truelove1997}. This pressure term is computed to ensure that the thermal Jeans length is resolved by at least a few cells \citep[four in our case, see][for details]{Teyssier2010} everywhere in the simulation, avoiding artificial fragmentation. This translates into the introduction of a lower limit in the EOS, thanks to a polytrope $T\propto \rho$, the so-called Jeans polytrope. Given our spatial resolution $\cellmin = 0.05 \pc$, this floor becomes active for densities above $\approx 3075\cc \times (\cellmin/1\pc)^{-4/3} = 1.7\times 10^5 \cc$, corresponding to a temperature of $\approx 101 \U{K} \times (\cellmin/1 \pc)^{2/3} = 14 \U{K}$ and a Jeans mass of $\approx 4250 \Msun \times (\cellmin/1 \pc)^{5/3} = 29 \Msun$. For the very low density regions (in the stellar halo), the EOS is a polytrope of index $5/3$ corresponding to the hot virialised gas. The EOS is applied in all the cells of the simulation, except in active \hii regions (see  Section~\ref{sec:fb}).

\subsection{Star formation}
\label{sec:sf}

According to Jeans' formalism, gaseous clumps more massive than the Jeans mass are unstable and should collapse and form stars. However, this is not the case in the simulation for two reasons: (1) our very low Jeans mass ($\sim 30 \Msun$) would represent an enormous amount of stellar particles to form and evolve, which would considerably slow down the computation, and (2) converting clumps of gas down to a few $10 \Msun$ into stellar particles of $\sim 0.5 \Msun$ means that individual stars would be resolved, which implies a proper treatment of the associated physics. First, the stellar initial mass function (IMF) should be sampled correctly, which is not feasible using a sub-grid recipe for star formation (see below). Second, two-body encounters, binary/multiple star formation should be accounted for via a collisional approach which is numerically extremely costly over an entire galaxy\footnote{Galaxy simulations generally assume that the stellar ``fluid'' is collision-less, to allow some approximations in the computation of the gravitational force by the way of numerical softening. This is not the case in dense systems where star-star interactions are frequent, i.e. where the relaxation time is short with respect to the lifetime of the system (e.g. in star clusters).}. Although such approaches are commonly followed thanks to dedicated numerical techniques in the context of molecular clouds and star clusters \citep{Aarseth2003}, the integration of the galactic scale effects in such simulations has started out only recently \citep{Renaud2011, Bonnell2013} and a fully consistent treatment over the entire galaxy is yet out-of-reach.

Therefore, the conversion of gas into stellar particles is artificially forced to occur at a lower \emph{mass} resolution, as follows. The local star formation rate (SFR) follows the \citet{Schmidt1959} law: $\rho_\mathrm{SFR} = \epsilon \rho/\tff \propto \epsilon \rho^{3/2}$, where $\epsilon$ is the star formation efficiency (SFE) and $\tff = \sqrt{3\pi/(32G\rho)}$ the free-fall time corresponding to the gas volume density $\rho$, but only when it exceeds a certain threshold. Physical interpretations of this threshold are discussed in \citet{Krumholz2012} and \citet{Renaud2012}. After tests, it appears that star formation in the simulation is very little sensitive to the density threshold, since the SFR for densities close to the threshold is very low and thus such gas does not significantly modify the global SFR. To speed-up the computation, we have set it to a high value ($2\times 10^3 \cc$), such that cells that would not, in any case, form a significant number of stars are not even considered in the numerical process. (This choice is further discussed and justified in Section~\ref{sec:pdf}.) Diffuse star formation might not be properly described, but this concerns an extremely small fraction of the global star formation, and should only affect the structure of the galaxy on very small scales ($\sim 1\mh 10 \pc$, $\sim 10 \Myr$), in the outer regions of the galactic disc.

At each timestep $\dd t$ ($\sim 10 \mh 100 \yr$, depending on the local level of refinement), in cells (of size $\cellsize$) denser than the threshold, a dimensionless number $n_\star$ is drawn, following a Poisson distribution of mean value $\rho_\mathrm{SFR} \cellsize^3 \dd t / M_\star$ \citep{Katz1992}:  a non-zero value of $n_\star$ implies the conversion of the mass of gas $n_\star M_\star$ into a stellar particle. We avoid creating solar-like mass stars by setting $M_\star = 160 \Msun$. This reduces the number of particles created, but at this resolution the mass of stars formed (and thus the SFR) is not affected by this artifact.

The SFE $\epsilon$ is chosen such that the global SFR over the entire galaxy matches the observations. This value is highly sensitive to the data and the method used: SFRs between $\sim 1 \Msun \yr^{-1}$ \citep{Robitaille2010} and $\sim 10 \Msun \yr^{-1}$ \citep{Guesten1982} have been proposed. We have adopted $\epsilon=3\%$, in agreement with \citet{Krumholz2007a} who suggested a universal value of a few percent, and measured a global SFR of $\sim 1 \mh 5 \Msun\yr^{-1}$, on average.

\subsection{Stellar feedback}
\label{sec:fb}

Our implementation of the stellar feedback encompasses three ingredients, all active in the first $10 \Myr$ of the life of the stars. Because we do not resolve individual stars, one stellar particle (of mass $M_\star$) represents a collection of stars of different masses. For simplicity, we only include the feedback effects from the massive end of the stellar mass function (i.e. the OB-type stars, $> 4 \Msun$, \citealt{Povich2012}), corresponding to $\etaob \approx 20\%$ of the mass of one stellar particle for a \citet{Salpeter1955} IMF; the other 80\% being feedback-inactive. In other words, for each of our stellar particle, the feedback is due to the mass $M_\star \etaob$. Other forms of feedback like stellar jets, gas recycling and an active galactic nucleus are not included.

\subsubsection{Photo-ionisation}

OB-type stars emit energetic ultraviolet photons that ionise the surrounding neutral gas, creating \hii regions. The size of an ionised region depends on the luminosity $L_\star$ of the central stellar source and the local opacity of the ISM through the density of electrons $n_e$ and the recombination rate $\rr$. To ensure an efficient numerical calculation, we have implemented a simple treatment of photo-ionisation (and radiative pressure, see below) in RAMSES: we define a \hii bubble as a \citet{Stroemgren1939} sphere centered on the source, of radius
\begin{equation}
\rhii = \left(\frac{3}{4\pi}\frac{L_\star}{n_e^2 \rr}\right)^{1/3}.
\end{equation}
The luminosity, in term of number of ionising photons, is a piecewise function of the age $a_\star$ of the source (Krumholtz, priv. comm.):
\begin{equation}
L_\star = L_0 M_\star \etaob \left\{\begin{array}{ll}
1& \textrm{for } \tff < a_\star \leq 4 \Myr\\
(4 \Myr) / a_\star & \textrm{for } 4 \Myr < a_\star < 10 \Myr\\
0 & \textrm{else},
\end{array}\right.
\end{equation}
where $L_0 = 6.3\times 10^{46} \U{s^{-1}\Msun^{-1}}$ and $\tff$ is the local free-fall time, used to delay the ignition of the photo-ionisation feedback after the formation of the first stellar particle, so that other stars are given the opportunity to form in the same region. The recombination rate is a function of the temperature of the bubble $\thii$, and reads $\rr = 2.1\times 10^{-10} \U{cm^3s^{-1}} (\thii/1\U{K})^{-3/4}$.

The positions and radii of the bubbles are updated at each timestep. Within the \hii bubble, the EOS defined in Section~\ref{sec:eos} is replaced by an isothermal branch: the gas temperature is set to $\thii = 2 \times 10^4 \U{K}$,  corresponding to a sound speed of $\sim 10 \kms$ \citep{Krumholz2009a}. When the cell containing the source of photo-ionisation is larger than the bubble (i.e. the \hii region is not resolved), the temperature of the gas in this cell is weighted by the volume ratio of the bubble and the cell. If a given cell comprises more than one unresolved bubble, they are all replaced with a unique bubble, conserving the total ionised volume. 

Eventually, two or more (resolved or not) \hii bubbles overlap. Since the overlap volume cannot be ionized twice, each bubble is grown such that the ionised volume is conserved, in a similar fashion as in \citet{Thomas2009}. This strategy ensures that the geometry of the ionisation front is globally preserved, even when bubbles form chains. However, when the overlap is a large fraction of the total volume of the bubbles (their separation is smaller than their radii), they are merged to speed-up the calculation, without a significant alteration of the geometry.

This method allows for a much faster treatment of the ionisation than the more accurate radiative transfer \citep[e.g.][]{Aubert2008}. Our approximation of Str\"omgren spheres does not allow us to retrieve fine structures in the ISM such as pillars, but the overall injection of energy is reproduced. A comparison with a more accurate treatment of the ionised regions using ray-tracing will be presented in a forthcoming publication (Chardin et al., in prep.).

\subsubsection{Radiative pressure}

Most of the momentum feedback is hauled by energetic, ionising photons. Therefore, using \hii bubbles as carriers of the momentum avoids the artificial definition of a mass loading factor. Inside each \hii bubble (after management of the overlaps), momentum is deposited as radial velocity kicks $\Delta v$ over the time interval $\Delta t$ matching the timestep of the simulation at the coarse level ($\sim 0.02 \Myr$):
\begin{equation}
\Delta v = \multiscat \frac{L_\star h\nu}{\mhii c}\Delta t,
\end{equation}
with $h$ the Planck constant, $c$ the speed of light, $\mhii$ the gas mass of the bubble affected by the kick and $\nu$ the frequency of the flux representative of the most energetic part of the spectrum of the source. For simplicity, we consider the luminosity of the Lymann-$\alpha$ photons and set $\nu=2.45\times 10^{15} \U{s^{-1}}$. The dimensionless number $\multiscat$ represents the average multiple scattering of photons as they travel through the bubble. It accounts both for multiple photon/electron collisions and the decay of the energy injected between each collision. We set $\multiscat=2.5$, similarly to \citet{Krumholz2010} and \citet{Dekel2013}.

Note that our velocity kicks do not have heuristic values and the radiative feedback is not related to the binding energy of the clump, contrarily to \citet{Oppenheimer2006}, \citet{Oppenheimer2010} or \citet{Hopkins2011}.

The delay in the ignition of the ionisation and radiative pressure represents a major limitation of the feedback recipe implemented. In real systems, the details on the formation of the first star in a cloud will determine the conditions for the formation of the next stars. A debate still exists on the exact role of radiative feedback in preventing, and/or triggering star formation, and in modifying the inner structure of clouds \citep[see][and references therein]{Dale2005, Dale2011, Tremblin2012}. The associated structures (pillars and globules) are created very early during the star forming epoch of a cloud and might strongly influence its future evolution. However, their size ($0.1 \pc$ and less) being very close to our resolution limit ($0.05 \pc$), the present simulation is inadequate to monitor their role and evolution. For these reasons, the feedback recipe has to be adjusted to the numerical resolution, which we do by implementing the delayed ignition.

\subsubsection{Supernova explosions}

The massive end of the IMF responsible for photo-ionisation feedback will explode as supernovae (SNe) after an average age of $10 \Myr$. The implementation of the supernova feedback, as a Sedov blast, is described in \citet{Dubois2008}. Because the gas in our simulation follows an EOS and not the contributions from heating and cooling terms (recall Section~\ref{sec:eos}), injecting thermal feedback would have no effect. Therefore, the total energy of the SN feedback ($10^{51}\erg$) is injected only in the kinetic form. The use of an EOS biases the thermodynamical evolution of the feedback energy injected by SNe. In particular, re-compression and re-heating in the reverse shock of SN blasts cannot be accurately followed using our prescription \citep{Chevalier1989}, which can affect the thermodynamical state of these regions.

\section{Initial setup and early formation of structures}
\label{sec:ic}

The general purpose of this simulation is to reproduce an isolated grand-design spiral galaxy and to focus on the structure of the interstellar medium at the highest resolution possible, and its evolution during a period long enough to monitor the coupling between stars and gas through feedback ($\sim 100 \Myr$). Cosmological aspects, the formation of the galaxy, the interaction with satellites and the long term evolution are not in the scope of this study.

\subsection{Initial conditions}
\label{sec:ic}

The initial conditions of the galactic model have been generated by {\tt pyMGE} (Python Multiple Gaussian Expansion, Emsellem \& Renaud, in prep.). This code uses the MGE method developed by \citet{Emsellem1994} to decompose the mass or luminosity density of the galaxy with a set of Gaussian functions \citep{Bendinelli1991, Monnet1992}. The gravitational potential is then rendered with particles and the velocities are attributed by solving the Jeans equations for each Gaussian. For the Milky Way, the input of {\tt pyMGE} is roughly based on the Besan\c{c}on model \citep{Robin2003}. We have used 31 Gaussians to render the spherically symmetric dark matter (DM) halo, the spheroid, the bulge, the thick and thin discs as well as a gas disc. The gas particles are used to ensure the global coherence of the model, so that the associated gravitational potential is accounted for, but they will be replaced by a grid-based description for the simulation itself.

\begin{figure}
\includegraphics{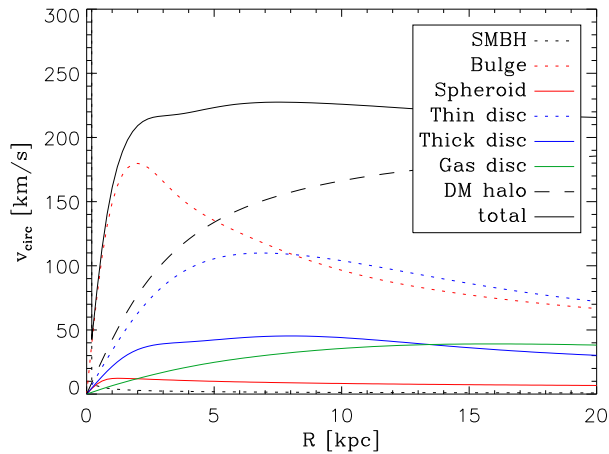}
\caption{Radial profiles of the circular velocity of the model components.}
\label{fig:vcirc}
\end{figure}

All components are initially feature-less and axisymmetric: the structures such as the bar and spirals will be created during the run, from instabilities in the velocity distributions according to the profiles chosen (see below). The non-gaseous components have been rendered using a total of $6\times 10^7$ particles. A point-mass super massive black hole (SMBH) has been initially placed at the center of the galaxy with a mass of $4\times10^6\ \msun$. The physical properties of the galaxy are summarized in Table~\ref{tab:init}. The velocity dispersions have been tuned during several attempts at low resolution to ensure the creation of spirals and a bar within a few ($\sim 2$) rotation periods. The rotation curves are plotted in \fig{vcirc}. The baryons represent 45\%, 24\% and 5\% of the mass at $8 \kpc$, $20 \kpc$ and large radius (i.e. if the halo would not have been truncated in the initial setup) respectively.

\begin{table}
\caption{Initial setup}
\label{tab:init} 
\begin{tabular}{lcccc}
\hline
Component & mass [$\times 10^9 \msun$] & number of particles\\
\hline
SMBH & $4.0 \times 10^{-3}$ & 1 \\
Bulge & 20.3 & $1.3 \times 10^7$ \\
Spheroid & 0.3 & $2  \times 10^5$ \\
Thin disc & 21.5 & $1.4  \times 10^7$ \\
Thick disc & 3.9 & $2.6  \times 10^6$ \\
Gaseous disc & 5.9 & ($\sim 2.4 \times 10^8 $ AMR cells)\\
Dark matter halo & 453.0 & $3.0  \times 10^7$ \\
Total & 500.0 & $6.0 \times 10^7$ \\
\hline
\end{tabular}
\end{table}

No satellite galaxy has been included in the simulation as they are expected to have only a mild impact on the morphology of disc (warp, precession), the structure of the ISM, or the star formation history during our time lapse of interest \citep{Bekki2012}.

\begin{figure*}
\includegraphics{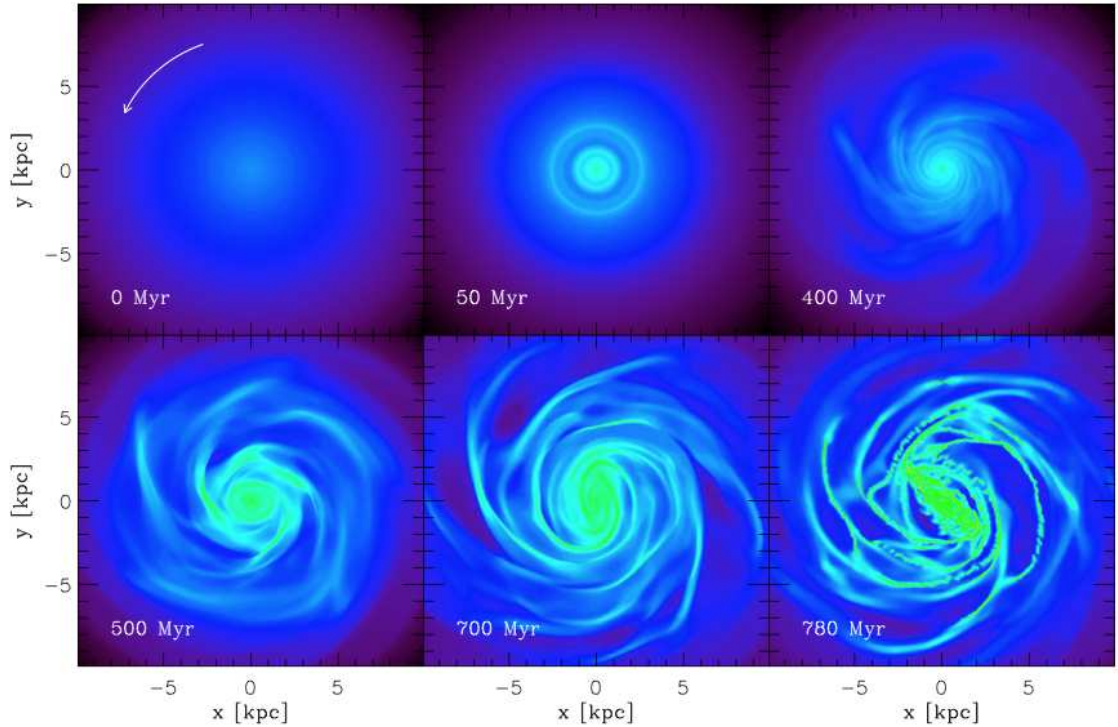}
\caption{Surface density of the gas disc seen face-on, at different epochs during the formation of structures from the initial setup (top-left). During this phase (up to $t \approx 750 \Myr$), the maximum resolution is set to $6 \pc$, to avoid the early fragmentation of the gaseous component into very small and dense clumps.}
\label{fig:evolution}
\end{figure*}

The gas disc is initialized on the AMR grid via an analytical profile. The radial- (respectively vertical-) profile follows an exponential distribution, with a scalelength of $6 \kpc$ (resp. $0.15 \kpc$) and is truncated at $28 \kpc$ (resp. $1.5 \kpc$). Beyond this truncation, the ``intergalactic'' density is $10^{-7}$ times that of the edge of galaxy. The initial gas mass is $5.94 \times 10^9 \Msun$, which represents $\approx 11\%$ of the total baryonic mass. No additional gas is accreted during the evolution of the model. The gaseous structures form during the simulation, following the instabilities of the stellar component (\fig{evolution}). The maximum resolution of the AMR grid has been progressively increased during this early phase (from $6 \pc$ to $0.05 \pc$), to avoid a too rapid fragmentation of the gas. Star formation has been activated once the structures had formed ($t > 730 \Myr$), preventing a premature consumption of the gas.

\begin{figure*}
\includegraphics{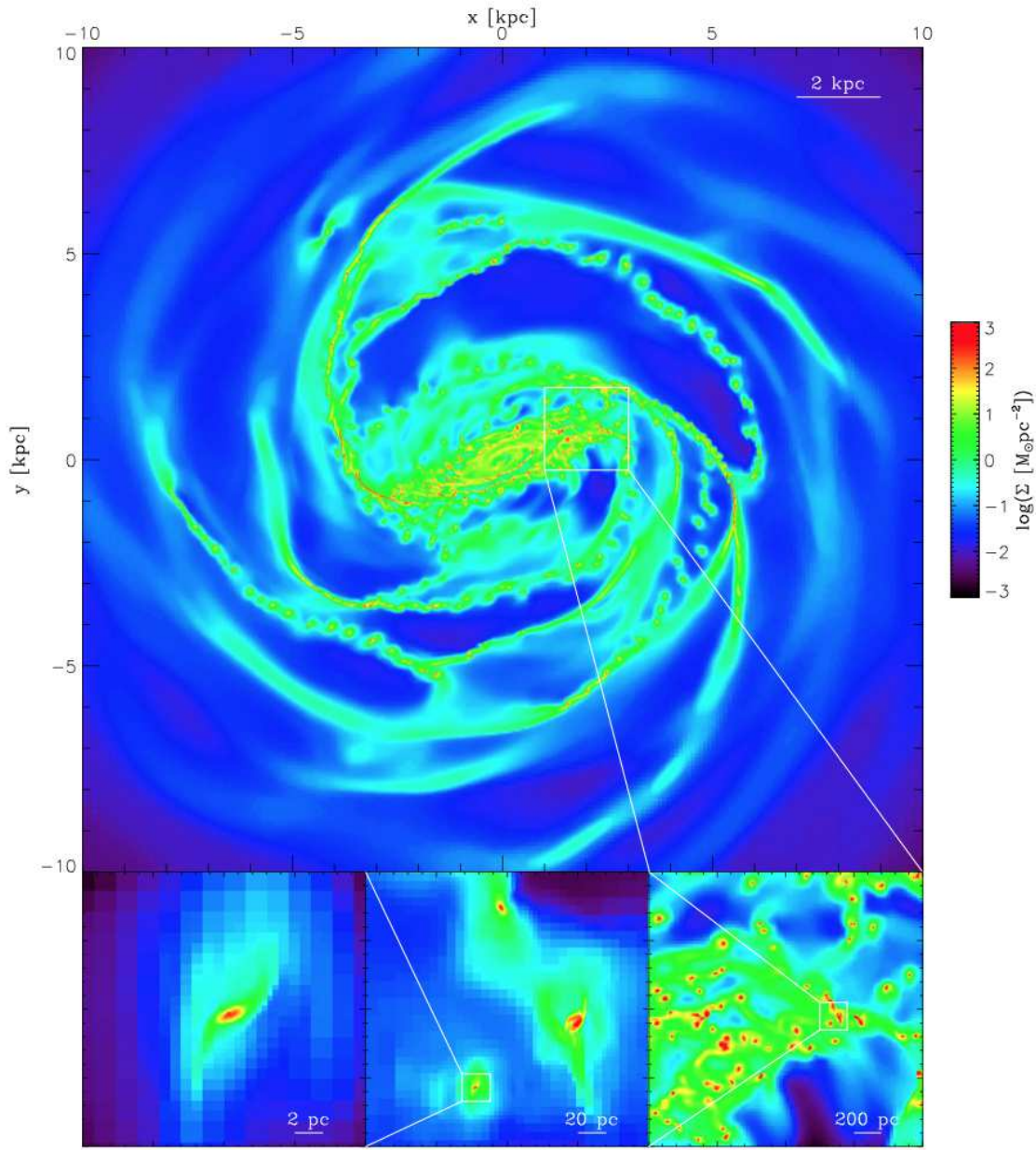}
\caption{Surface density of the gas disc seen face-on, $800 \Myr$ after initial setup. The bar and spiral arms host dense clumps, some of which being distributed as ``beads on a string''. The color table only applies to the main panel: the table has been changed in each zoom-in view to enhance contrast.}
\label{fig:t354}
\end{figure*}

\begin{figure}
\includegraphics{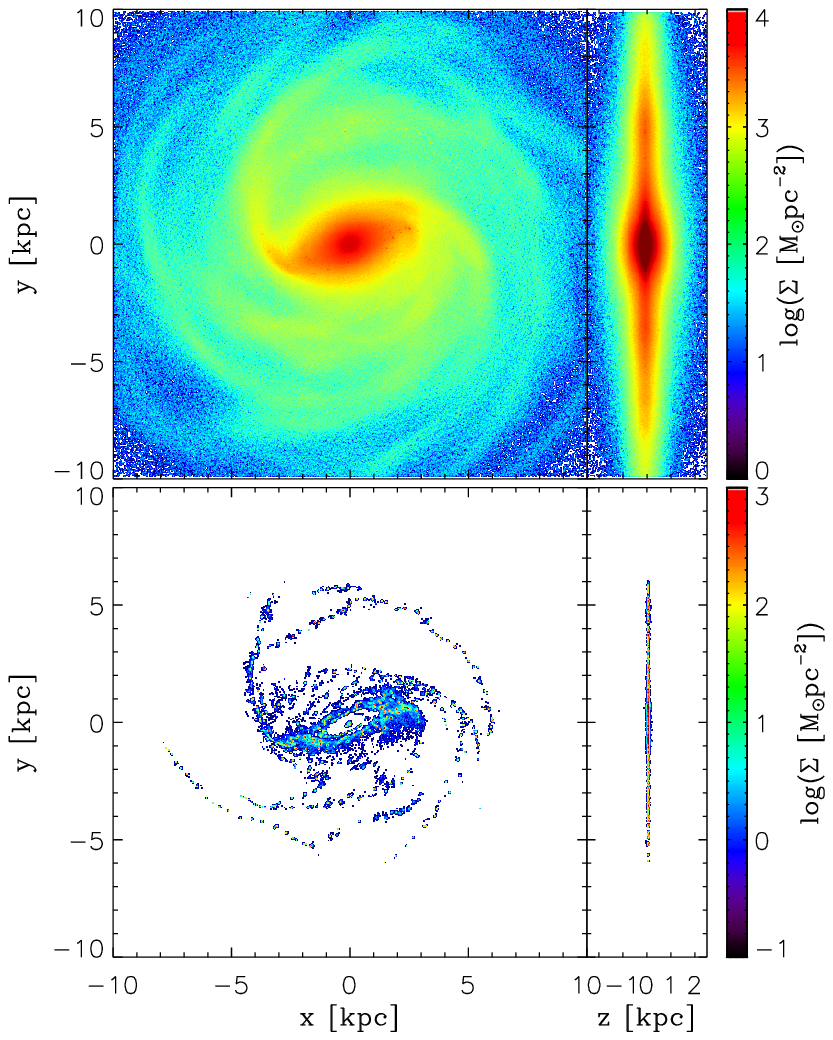}
\caption{Surface density of the primitive stars (top), and the stars formed during the simulation (bottom), $800 \Myr$ after initial setup.}
\label{fig:stars}
\end{figure}

\fig{t354} displays the surface density of the gas after most of the kpc-scale morphology (bar, spirals) and sub-parsec structures (in molecular clouds) have formed\footnote{An interactive version of this map, at full resolution, is available here: http://irfu.cea.fr/Pisp/florent.renaud/mw.php}. The bar and the spirals, formed from instabilities in the initial conditions, are also detected in the stellar component both primitive (i.e. particles set in the initial conditions) and newly-formed out from gas consumption, as shown in \fig{stars}. The properties of the young star clusters, and in particular the regularity of their separation along spiral arms, are discussed in Section~\ref{sec:boas}.

Interestingly enough, the galactic bar hosts a large region ($\sim 1 \kpc^2$) completely free of star formation: dense clouds are found at larger radii, along the border of the bar, and in the innermost $\sim 200 \pc$, but not in between. Although our high density threshold for star formation could cause such deficit, it is the absence of gas clouds, even at low-density, in this area that indicates the origin of the phenomenon. This particular feature will be discussed in more details in a forthcoming paper: the preliminary results indicates that the shear, and to a lower extent the tidal field, dissolve the gaseous clumps falling in this region. Then, a given packet of gas does not reach a high enough volume density over a long enough period of time for stars to form. At smaller radii however, the gas accumulates around the SMBH and does form stars (see the next Section).

\subsection{Bar, spirals, nuclear ring and resonances}
\label{sec:resonances}

Using the angular velocity of the stellar bar at $t \approx 800 \Myr$, we found a pattern speed of $\omegap = 58 \kms\kpc^{-1}$, in agreement with values from the literature about the real Milky Way \citep[$\omegap = 59\pm 5 \kms\kpc^{-1}$,][]{Debattista2002}, although a rather large uncertainty exists on local standard of rest data. The epicycle frequency $\kappa$ is computed numerically: $\kappa(R)^2 = R \dd \Omega^2 / \dd R + 4\Omega^2$, where $R$ represents the distance to the galactic center, in the plane of the disc \citep{Binney2008}.

\begin{figure}
\includegraphics{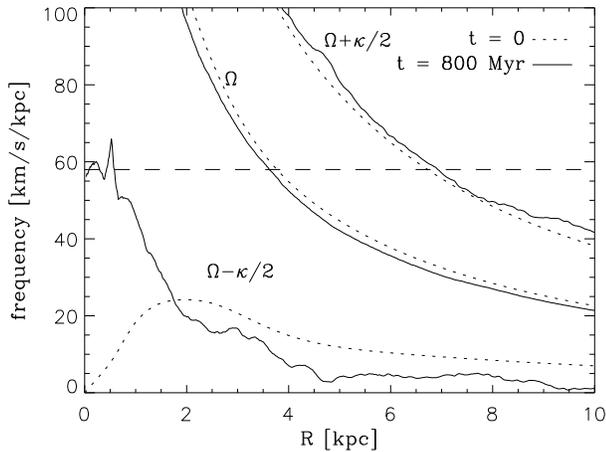}
\caption{Frequency diagram from the initial conditions (dotted curves) and at $t \approx 800 \Myr$ (solid lines). The measured pattern speed of the bar ($\omegap = 58 \kms\kpc^{-1}$, horizontal line) gives a corotation at $3.6 \kpc$, and an outer Lindblad resonance at $6.3 \kpc$. Inner Lindblad resonances appear in the inner kpc during the formation of the bar (see also \fig{ilr}). (The $\Omega \pm \kappa/2$ curves have been low-pass filtered to reduce the noise.)}
\label{fig:epicycle}
\end{figure}

The frequency diagram on \fig{epicycle} shows the resonances found in the stellar discs, from the initial conditions ($t=0$) and at $t = 800 \Myr$. The evolution of the galaxy and the formation of structures (the bar in particular) modifies the frequencies in the inner $\sim 2 \kpc$, but the outermost regions remain relatively unchanged, because less massive structures (e.g. spiral arms) form there. The corotation ($\Omega = \omegap$) corresponds to a radius of $3.6 \kpc$, i.e. about $1 \kpc$ further than the tip of the bar. The outer Lindblad resonance (OLR, $\Omega + \kappa/2 = \omegap$) is found at $6.3 \kpc$, i.e. close to the ``knee'' visible in several gaseous arms in \fig{t354}.

\begin{figure}
\begin{center}
\includegraphics{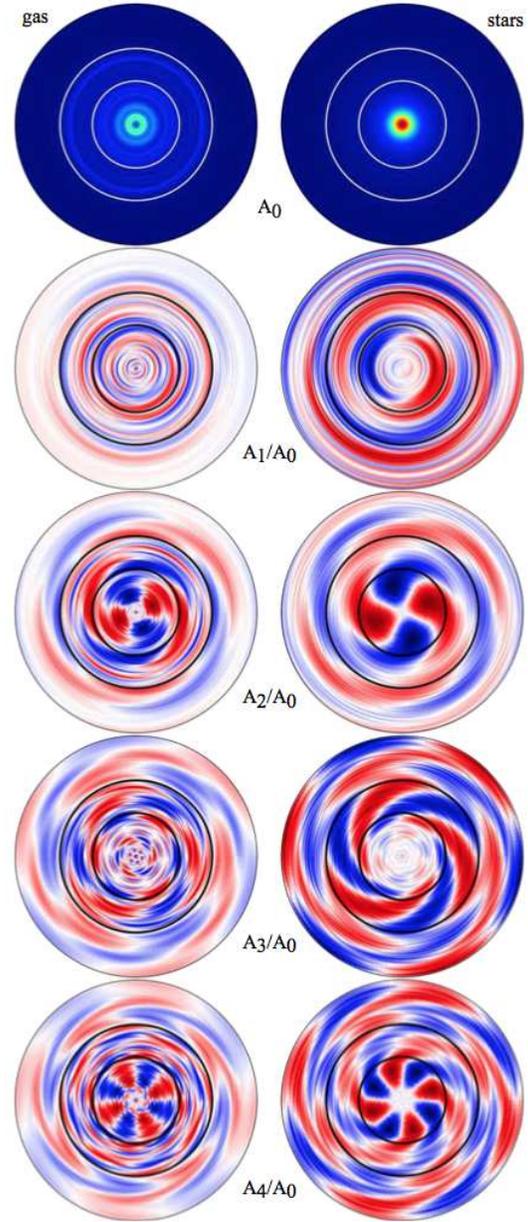}
\end{center}
\caption{Amplitudes of the fundamental and the first four modes (normalised to that of $m=0$) from the Fourier transform of the surface density of the gaseous (left) and stellar (right) discs, in the innermost $10 \kpc$, at $t=800 \Myr$. The two circles indicate the corotation ($3.6 \kpc$) and the OLR ($6.3 \kpc$), as estimated in \fig{epicycle}. For $m>0$, the blue to red colours label negative to positive amplitudes. The color bar varies for each panel, to enhance contraste.}
\label{fig:fft}
\end{figure}

As a complement, \fig{fft} shows the amplitude of the first $m$-modes of the Fourier transform of the surface density of the disc $\Sigma$ in polar coordinates $(R,\theta)$
\begin{equation}
\Sigma(R,\theta) = \sum_{m=0}^{\infty} A_m(R) \cos{\left[m\theta - \Phi_m(R)\right]},
\end{equation}
as in \citet{Kraljic2012}, for both the gaseous and the stellar components. The highest amplitudes in the mode $m=2$ correspond to the bar and the symmetric pair of arms, in the inner $\sim 4 \mh 5 \kpc$. The distinction between the two is mainly seen as a change of the phase at about the corotation radius, but also as a drop of the amplitude $A_4/A_0$ at the end of the bar. As visible in \fig{t354} and \fig{stars}, the $m=2$ symmetry is progressively replaced with a collection of secondary gaseous arms and a more uniform distribution of stars, i.e. a growing importance of high-$m$ modes with increasing radius. The amplitude of the even $m$ modes ($A_0$, $A_2$, $A_4$) drops at about $8 \kpc$, in both components, revealing the radial extension of the spirals. Although gas spirals are less sensitive to truncation of dynamical origin, the gaseous spirals do not extend further than their stellar counterpart. 

Both stellar and gaseous spirals extend further than the OLR found using the pattern speed of the bar, which suggests that other, secondary, pattern speeds may exist in the disc, modifying the first order picture drawn above. In particular, we found a slower pattern speed for the spirals ($\approx 50 \kms\kpc^{-1}$) than that of the bar ($\approx 58 \kms\kpc^{-1}$). Using this additional value on the frequency diagram of \fig{epicycle} leads to a secondary OLR at $8 \kpc$, i.e. at the detected truncation of the spirals. Although it is subject to uncertainties, this confirms the idea of \citet{SellWood1988} on the role of secondary pattern speeds. However, the radial extension of the spirals is still too small to explain the GALEX observations of ultraviolet light tracing star formation in the outer galactic discs in the local Universe (see \citealt{Thilker2007}, see also \citealt{Barnes2012} for optical data). Either such structures might rather be of tidal origin following an interaction with another galaxy \citep{Thilker2007}, or because self-gravity of the arm is strong enough to overtake the influence of the resonance and form un-truncated structures, or more simply because new arms could have form \emph{in situ}, independently of structures at smaller radii. Also the coupling with a flatten dark matter halo might allow for an expansion of star forming gas beyond $10 \mh 15 \kpc$.

Additional pattern speeds might develop with the formation of inner structures and would lead to other resonances, but their transitory aspect and the relatively low mass involved makes them less relevant for the large-scale study presented here.

\begin{figure}
\includegraphics{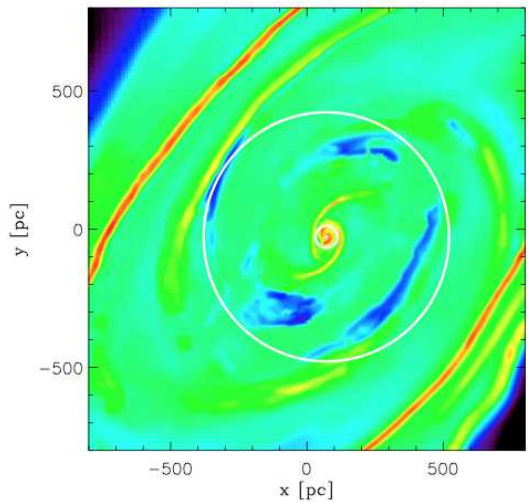}
\caption{Inner structures of the gas disc (displayed as surface density). A nuclear disc has formed in the central $\sim 30 \pc$. The cross marks the position of the SMBH and the 2 circles correspond to the inner Lindblad resonances detected at $\sim 40 \pc$ and $\sim 450 \pc$, when they are the most visible (at $t = 750 \Myr$).}
\label{fig:ilr}
\end{figure}

Once the bar is formed, inner Lindblad resonances (ILRs, $\Omega-\kappa/2 = \omegap$) can be found in the innermost kiloparsec using the pattern speed proposed above, while it was not the case from the initial conditions. However, the uncertainties on this pattern speed, the possible existence of others and the noise introduced by our method in the estimation of $\Omega-\kappa/2$ makes it difficult to determine precisely the radii of these ILRs. By evaluating the circular velocity from the mass profile, we found two ILRs at $\sim 40 \pc$ and $\sim 450 \pc$, as shown in \fig{ilr}. Both ILRs mark the innermost radii of spiral-like structures, at the edge of a nuclear disc and inside the galactic bar, respectively. The formation of the nuclear disc, its evolution and the fueling of the SMBH will be covered in a forthcoming paper.

\section{The ISM at large-scale}
\label{sec:largescale}

\subsection{Longitude-velocity map}

\begin{figure}
\includegraphics{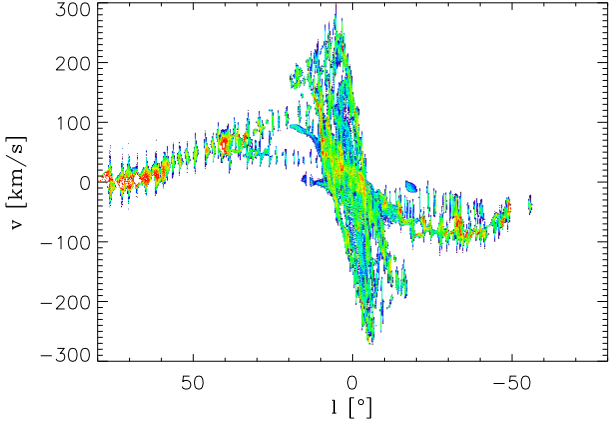}
\caption{Longitude-velocity map of the gaseous component at $t = 800 \Myr$. The color is a proxy for the CO flux in arbitrary units (see text).}
\label{fig:posvel}
\end{figure}

\fig{posvel} shows the radial velocity as a function of the galactic longitude, both evaluated from the approximate position of the Sun ($x = -8.5 \kpc$, $y = 0 \kpc$ in \fig{t354}, leading to a longitude of $23^\circ$ for the nearest tip of the bar). The color is a proxy for the CO flux along each line of sight, at the resolutions of $\approx 0.3^{\circ}$ and $\approx 1 \kms$ for the longitude and velocity respectively. In each gas cell of density between $25 \cc$ and $3000 \cc$, the flux is approximated by the gas mass divided by the square of the distance of the cell to the Sun.

The overall structure of this map matches well the observations in CO of \citet{Dame2001}: the position and velocity range of the bar, as well as the structure of the arms is well retrieved. The bright structure at $\sim 75^\circ$ corresponds to a nearby spiral arm, visible at $x \approx -8 \kpc$ and $y \approx -2 \kpc$ in \fig{t354}, i.e. the equivalent of the Sagittarius arm. Note that our simulation does not reproduce the nuclear ring in the inner $200 \pc$. The dynamical evolution of the inner bar might create such structure at a later stage.

Despite a broad selection in density, our map reveals several gaps between ``emitting clouds'', which is linked to the beads-on-a-string morphology discussed below and indicates both the absence of light diffusion in our simple post-process and a lack of extended, inter-cloud, dense gas. We note that our feedback recipe does not significantly spread the dense clouds into a more diffuse ISM.

\subsection{Gas density PDF}
\label{sec:pdf}

\begin{figure}
\includegraphics{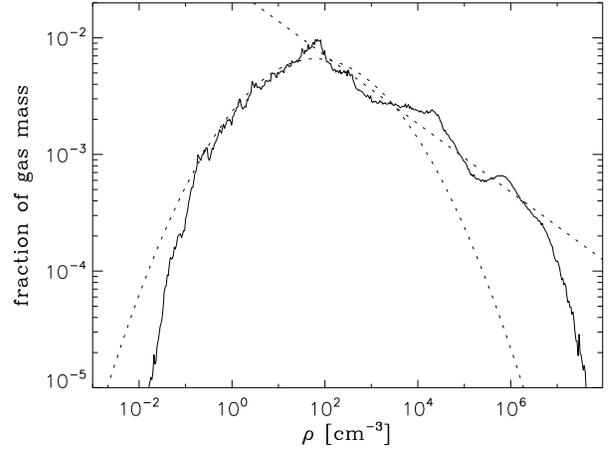}
\caption{Mass-weighted density PDF of the gas, and the best fit using a log-normal functional form, and a power-law (dotted lines).}
\label{fig:pdf354}
\end{figure}

The density probability distribution function (PDF) indicates in which state the gas is distributed over the galaxy, between the diffuse ISM and the dense clouds, and tells us about the relative role of gravity and turbulence in shaping the density field of the gas. The mass-weighted PDF of the entire galaxy, shown in \fig{pdf354}, can be approximated by a log-normal functional form, to first order, for the non self-gravitating turbulent gas. Log-normals generally provide good fits to PDFs of simulated isothermal, supersonically turbulent gas \citep[among many others, see][]{Vazques1994, Nordlund1999, Wada2001}. However, the resolution reached allows us to probe the very dense gas of molecular clouds (i.e. $\gtrsim 10^{5\mh 7} \cc$), which appears to be in excess with respect to a log-normal fit, by a factor up to $\sim 10$. In this regime, the PDF develops a power-law tail, as shown in \fig{pdf354} and already noted by several authors \citep[e.g.][]{Klessen2000, Hennebelle2008, Vazques2008, Audit2010}. \citet{Elmegreen2011} explained that such a shape results from the convolution of the ``classical'' log-normal form for the turbulent non-self-gravitating gas, with the PDF of the self-gravitating clouds, for which the radial density profile goes as $\rho \propto r^{-\beta}$. The resulting PDF would then exhibit a power-law tail of index $3/\beta - 1$, as shown in \fig{pdf354}. The best fit of our measured PDF in the range $50 \cc < \rho < 10^6 \cc$ gives out a slope of $\approx 0.29$, hence $\beta \approx 2.33$. We found a comparable index $\beta$ in the radial density profile of the gas clumps of the simulation in this density range. We note it is also comparable to the analytical solution of a self-similarly collapsing isothermal sphere \citep[$\beta = 2$,][]{Shu1977}. \citet{Lombardi2008, Lombardi2010} noticed a similar excess of dense gas with respect to a log-normal distribution in the observations of Milky Way star forming clouds, via measures of columns density extinctions.

The power-law tail of the PDF diverges from the log-normal at about the density threshold used for star formation (Section~\ref{sec:sf}). This coincidence is neither a cause nor a consequence of our choice for threshold: 
the process of star formation tends to empty the reservoir of gas at densities above the threshold, which goes in the opposite direction than what is measured here. Our analysis suggests to opt for such a threshold in order to ensure that simulations create stars in self-gravitating gas. Because of technical limitations, most of the previous numerical studies did not reach the resolution needed to probe the very dense gas, and thus to detect the onset of the regime in which self-gravity dominates over turbulent support, or in other words, the transition from the log-normal to the power-law tail. In this case, the threshold must be chosen from other physical considerations at larger scales, like the self-shielding of clouds to external radiation ($\sim 0.1 \mh 1 \cc$, \citealt{Schaye2004}) or the formation of molecules ($\sim 100 \cc$, \citealt{Krumholz2009a}). For example, while assuming a log-normal PDF, \citet{Renaud2012} proposed a density threshold corresponding to the onset of supersonic turbulence ($\sim 10 \cc$), leading to shocks and gas compression. Presumably, at ``higher resolution'', i.e. when considering smaller scale physics, the PDF would yield a power-law tail and the threshold should be set to a higher value. Note that the onset of self-gravity might not occur at an universal density, depending, among other things, on the turbulence.

In the PDF of \fig{pdf354}, at very low density ($\lesssim 10^{-1} \cc$), turbulence is not supersonic and thus the log-normal form does not match the data, while at very high density ($\gtrsim 10^7 \cc$), the effect of finite resolution of the simulation biases the results and the PDF in this regime probably results from a numerical artifact.

The alternative functional form of the PDF proposed by \citet{Hopkins2013} does not lead to a better fit of our data since its deviation from the log-normal (his parameter $T$) cannot model the off-peak excess of dense gas (i.e. the power-law tail) detected here.

\begin{figure}
\includegraphics{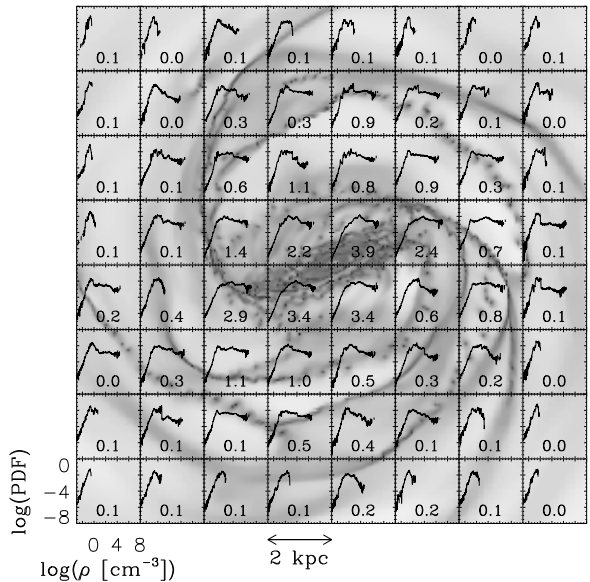}
\caption{Mass-weighted PDF of the gas computed on $2 \kpc \times 2 \kpc$ tiles. In each tile, the number indicates the Mach number.}
\label{fig:pdfmosaic}
\end{figure}

PDFs computed in smaller regions are plotted in \fig{pdfmosaic}, revealing the importance of the power-law regime with respect to the log-normal shape when self-gravitating structures exist in a tile. The Mach numbers $\mach$, indicated in each tile of \fig{pdfmosaic}, represent the mass-weighted average of the Mach numbers computed via the one-dimension velocity dispersion and the mass-weighted temperature at the scale of $1 \pc$. They can be connected to the width $\sigmapdf$ of the log-normal PDF through $\sigmapdf^2 = \ln{\left(1+b^2\mach^2\right)}$, with $b$ describing the relative importance of solenoidal and compressible modes of turbulence \citep{Federrath2008, Federrath2010, Molina2012}. However, the presence of the power-law tail voids this formalism, such that highly supersonic regions might correspond to not so extended PDFs and \emph{vice versa}.

\subsection{Beads on a string}
\label{sec:boas}

\begin{figure}
\begin{center}
\includegraphics{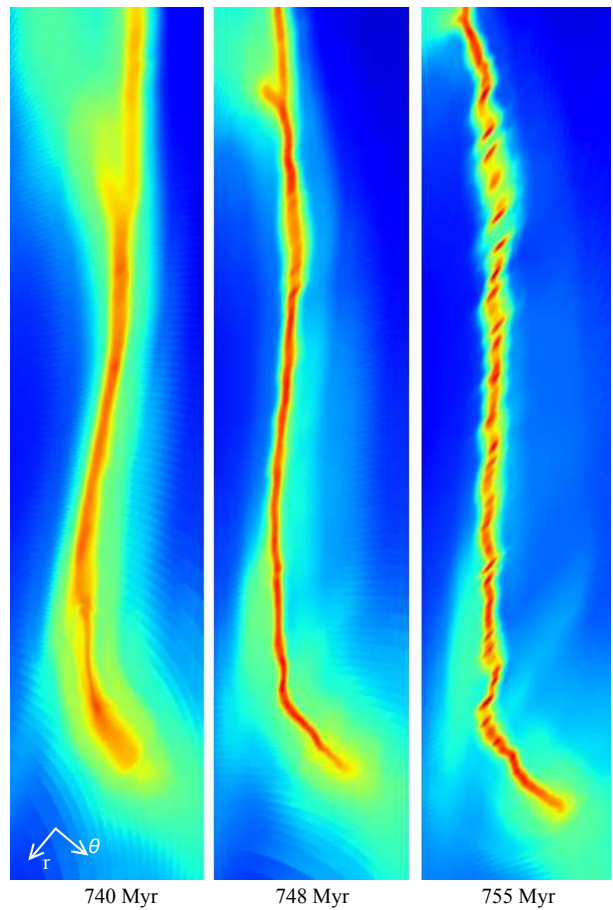}
\end{center}
\caption{Formation sequence of the beads on a string along one spiral arm. The arm has been projected into polar coordinates, for display purposes only (the orientation is indicated on the leftmost panel). The creation of dense seeds begins in the central panel and goes on under the effect of self-gravity in the right panel. A later stage of the evolution is shown in \fig{feedback_clumps}.}
\label{fig:beads}
\end{figure}

Along the spiral arms, overdensity clumps have formed in a regular structure, like \emph{beads on a string} with a relatively uniform separation. The evolutionary stage of the beads on a string varies from arm to arm, such that an almost complete sequence can be seen in the single snapshot of \fig{t354}. Clouds connected by a faint, diffuse arm (e.g. in the outermost spiral arm in the bottom-left corner in \fig{t354}, and \fig{feedback_clumps}) have formed early, while regular long and thin arms showing a low density gradient along themselves (e.g. the arm connected to the left tip of the bar) still host the formation of clumps around dense seeds. For this reason, defining a formation time of the beads is somewhat arbitrary. However, the morphology of the host spiral goes from a smooth and continuous structure to a chain of overdensities connected by an arm $\sim 200$ times less dense in $10 \mh 15 \Myr$ (estimated by eye), as illustrated in \fig{beads}. This time lapse, interpreted as a free-fall time, corresponds to the density at the edge of the fragmenting spiral and also to that of the interclump medium once the beads have formed ($\approx 30 \cc$). The clumps themselves, although denser, do not immediately fragment because of pressure support.

The regularity in the separation of the overdensities is related to the process of their formation and further evolves with the host spiral. Although clumps form through gravitational collapse, they do so in elongated arms, i.e. not in an isotropic, three-dimensional structure. Therefore, the three- and two-dimensional Jeans' formalisms do not apply and one cannot link the separation of the clumps to the local Jeans length. \citet{Elmegreen1983} studied the relation between the separation of the beads and the width of their host spiral, in several nearby galaxies. They found that the ratio of the two is constant across the galaxy, and from galaxy to galaxy, independently of their Hubble type.

At much smaller scale, \citet{Fischera2012} studied the separation of clumps along interstellar filaments and found a relation with the pressure excess (in the clouds compared to the edge of the filament). For a given filament width, a high overpressure implies a shorter separation. The EOS used in our simulation (recall \fig{eos}) forbids the coexistence of a warm and a cold phase at a given density, contrarily to other models of the ISM which allow it \citep[see e.g.][]{Audit2010}. Therefore, the simulation lacks pressure from a hot diffuse gas to confine the spirals and the beads. However, and despite a large spread in the relation, we note that the detection of a high overpressure ($> 10$) corresponds indeed to smaller values of the separation/width ratio, as predicted by \citet{Fischera2012}. 

\begin{table}
\caption{Beads on a string in the outermost spiral}
\label{tab:boas} 
\begin{tabular}{cccc}
\hline
time [Myr] & mass [$\log (\msun)$] & radius [pc] & separation [pc]\\
\hline
780 & 6.2 & 168 & 270 \\
800 & 6.0 & 149 & 380 \\
820 & 5.5 & 82 & 430 \\
\hline
\end{tabular}
\end{table}

Table~\ref{tab:boas} displays the evolution between $t=780 \Myr$ and $t=820 \Myr$ of the properties of the gas clumps (denser than $500 \cc$) selected from the beads on a string structure of \fig{beads} and \ref{fig:feedback_clumps}. Within these $40 \Myr$, some clouds merge forming a more massive and larger cloud, but despite this, the average separation between the clumps increases with time, while their mass and radius decreases, as also visible on the global scale in \fig{clump_mf}. While the \emph{time evolution} of the separation is at least partially linked to the differential rotation of the disc (and thus the elongation of the spiral), the evolution of the mass and radius seems rather independent of the environment and only connected to internal processes (collapse, star formation, feedback). Estimates of the SFR in individual clumps ($\sim 2\times10^5 \Msun$ formed over $20 \Myr$ in the region mentioned above) indicate that the conversion of gas into stars is the main factor of the decay of the gas mass. The depletion of gas and further (re-)collapse of the cloud lead to the decrease in radius. Note however that the space and time resolutions of this simulation do not allow us to follow a complete life cycle of the clouds (formation, collapse, destruction, re-formation, etc). At a later stage, the dynamical evolution of the gaseous spirals and recycling of the molecular material will probably modify the appearance of the structures, leading to wider spirals and altering the regularity in the spacing of the clumps, as seen in recent interferometric observations of M~51 by \citet{Schinnerer2013}. Such evolution is also likely to be connected with a smoothing of the arm to inter-arm density contrast.

\subsection{Spurs}

\begin{figure*}
\begin{center}
\includegraphics{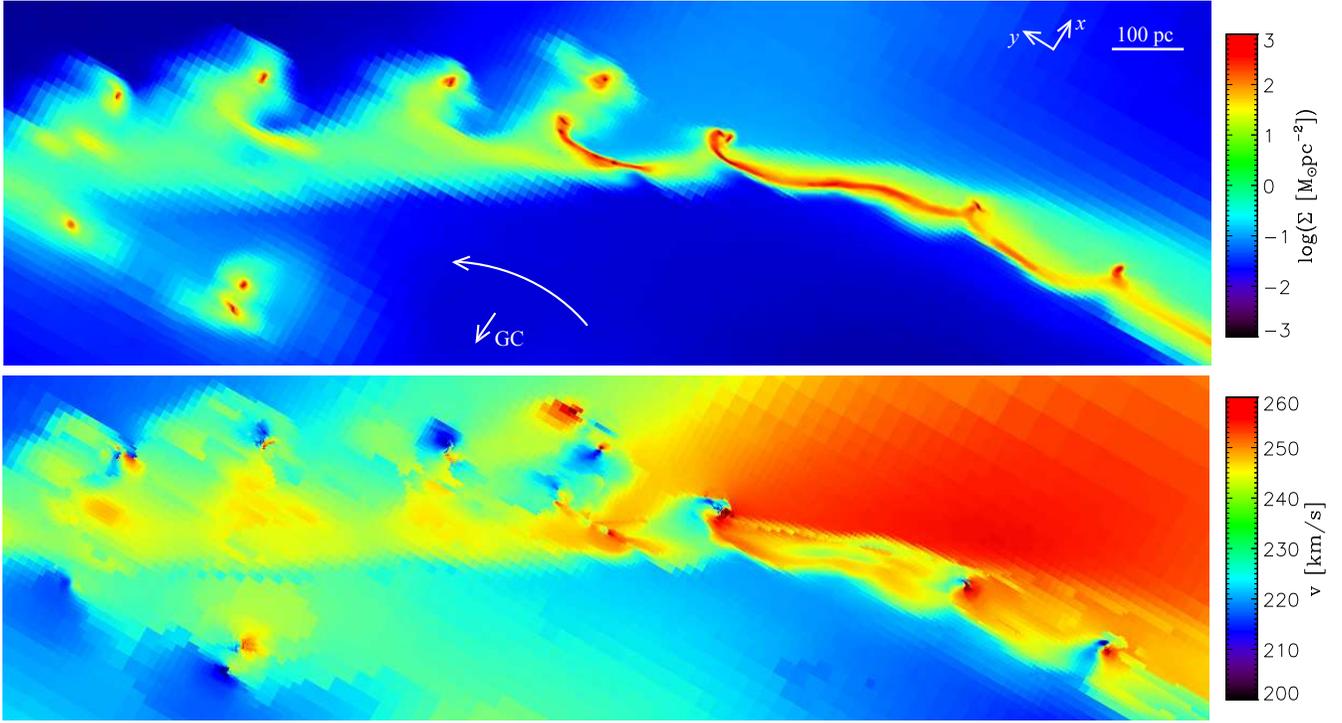}
\end{center}
\caption{\emph{Top}: surface density of gas in a spiral arm of our model of the Milky Way (at about $5 \kpc$ of the Galactic Centre, at $t=800\Myr$). The densest features are organised as spurs along the arm, resembling the pattern created by Kelvin-Helmholtz instabilities. The coordinate system shown is that of \fig{t354}, the small arrow points towards the Galactic Centre (GC) and the big one indicates the rotation of the disc. \emph{Bottom}: velocity along the horizontal axis of the figure, in the galactic reference frame. The strong velocity gradient (right half of the figure) generates Kelvin-Helmholtz instabilities that evolve into spurs (as in the left half).}
\label{fig:spurs}
\end{figure*}

Apart from the beads on a string, some gaseous spirals host another pattern of dense structures, as shown in \fig{spurs}. On the right half of the figure, a velocity difference of $\approx 40 \kms$ between both sides of the arm is associated with the thinnest and densest spurs, whereas on the left half, similar but more diffuse and thicker structures exist, with a weak velocity gradient through the arm. The later formed $\approx 15 \Myr$ ago, while the former are younger ($\approx 10 \Myr$). More generally, the sharpness of the density profile and the thickness of the spurs indicate an age gradient running from old structure on the left to younger, still forming spurs on the right. All these spurs ressemble Kelvin-Helmholtz instabilities as commonly observed at the interface between two fluids with a velocity difference.

The question of the physical driver of the formation of spurs has been addressed in many numerical works. \citet{Chakrabarti2003} detected them in two-dimensional, purely hydrodynamics models and explained their formation and growth invoking resonances and self-gravity. Without self-gravity, \citet{Wada2004} proposed a Kelvin-Helmholtz origin. Later, \citet{Kim2006} noted that a very strong spiral potential was necessary to trigger Kelvin-Helmholtz instabilities in two-dimensional models. By including magneto-hydrodynamics (MHD), they created spurs through magneto-Jeans instability in a three-dimensional description. \citet{Shetty2006} emphasized the role of MHD in creating and maintaining spur structure in grand-design spirals. \citet{Dobbs2006} presented simulations of analytical spiral potential with arbitrary pattern speed, including the formation of spurs at relatively low temperature ($\lesssim 1000 \U{K}$), as an evolution of clumps formed in the arms. The low Reynolds number of their model indicated that their structures did not form from Kelvin-Helmholtz instabilities. Our spurs differ from those detected by \citeauthor{Dobbs2006}: their connection with the spiral dissolves faster, they are shorter and host a dense clump at their tips. In that sense, the structures seen by \citeauthor{Dobbs2006} might rather be ``feathers'', as opposed to more compact spurs we and others have identified \citep[e.g.][]{Wada2004}. Note that these differences of geometry and density impact on off-spiral star formation: feathers host no or very little star formation while spurs form stars at a comparable amount and rate than other types of clouds.

The regularity in the spatial distribution of spurs is reminiscent of that of beads on a string. However, the formation process of the latter (fragmentation of an elongated structure, see Section~\ref{sec:boas}) clearly differs from that of the formers (velocity shear). Another difference lies in the position of the overdensity with respect to the spiral: while beads are centered over the width of their host gaseous arm, spurs are systematically offset toward the most rapid side of the spiral. Therefore, most of the clustered star formation in these regions does not occur inside the gas spiral but on its convex side, in dense, gravitationally bound clouds found next to the tip of each spur. \citet{Schinnerer2013} have recently detected a similar geometry in their comparison of CO(1-0) emission to near ultraviolet maps in M~51. They found that the young stars are offset by $\sim 100 \pc$ on the leading side of spiral arms (see their figure 8). In fact, spurs are common features of grand-design spiral galaxies, as noted by e.g. \citet{Elmegreen1980}, \citet{LaVigne2006} and \citet{SilvaVilla2012}.

Independently of the numerical codes and methods used, all previous simulation works mentioning spurs and/or feathers showed a rather ubiquitous distribution along the spirals \citep{Wada2004, Shetty2006, Khoperskov2013}. These studies set up analytical, well defined spiral potentials, reproducing the spiral strength seen in grand-design galaxies. In our case, spur-like structures only exist in strong continuous arms (grand-design-like), as already suggested by \citet{Shetty2006}. In the arm shown in \fig{spurs}, the spur pattern extends up to the point where the gas spiral splits in a Y-shape (not visible in this figure, but see \fig{t354} at $x\approx 5.5 \kpc$, $y\approx -2 \kpc$), marking the transition from a strong $m=2$ mode to higher order modes (see Section~\ref{sec:resonances}). Where higher orders develop (at larger radii along the spirals), the existence of spurs seems to be replaced by that of beads on a string.

\subsection{Asymmetric drift and feedback}
\label{sec:drift}

\begin{figure*}
\begin{center}
\includegraphics{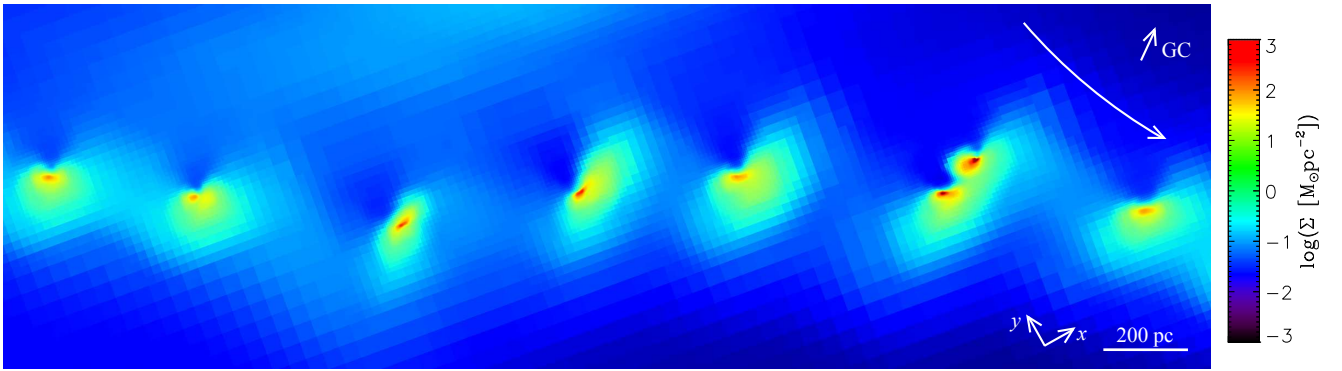}
\end{center}
\caption{Evolved beads on a string clouds in the outer spiral arm (at about $7 \kpc$ of the Galactic Centre, at $t=800\Myr$). The contrast of volume density between the beads and the string has reached $10^4$. As the stars decoupled from their gaseous nursery, supernovae explode away from the center of the clouds, which creates an asymmetric conic-like structure of low density (opened toward the top of this image). The clouds themselves are very mildly affected by the SN feedback. The coordinate system shown is that of \fig{t354}. The small arrow points towards the Galactic centre, and the big one indicates the rotation of the disc.}
\label{fig:feedback_clumps}
\end{figure*}

\fig{feedback_clumps} zooms-in on the same beads on a string as in \fig{beads} at a late stage of their evolution, once the chain connecting them has almost dissolved. At the time of their formation, the stars share the velocity dispersion $\vdisp$ of their gas nursery ($\approx 10 \kms$). While the cloud continues to collapse, the velocity dispersion of both the gas and the stars keeps increasing. On the one hand, for the stars, this is added to the ``intrinsic'' increase due to relaxation (which is underestimated in our numerical treatment by the use of a particle-mesh integrator). On the other hand, the gas is dissipative: its velocity dispersion increases slower than that of the stars, and may even decrease. The stellar and gaseous components decouple. After $10 \Myr$, we measure dispersions of $\approx 15 \kms$ for the young stars and $\approx 9 \kms$ for the gas. The rotation velocity being a combination of the circular velocity $\vcirc$ (almost constant for both components) and of the velocity dispersion as $\vrot^2 \approx \vcirc^2 - \vdisp^2$, the stars drift away from the gas cloud, at about $0.4 \kms$ (the so-called asymmetric drift, \citealt{Binney2008}). When the supernovae explode, they lag a few parsecs behind the densest part of the cloud in which they were born. The blasts occur in a medium showing a density gradient, and thus expand faster away from the cloud than toward its centre, and this effect amplifies as new SNe, continuously formed by the remaining cloud, explode in an already affected medium. The result is visible in \fig{feedback_clumps}, as cones of low density behind each cloud. Since the separation between clumps is several $100 \pc$ and the intra-cloud medium is diffuse ($10^{-3}\mh 10^{-2} \cc$) \emph{in this region of the Galaxy}, stellar feedback does not create dense spherical shells which could further fragment \citep{Norman1980} and form a secondary generation of stars out of an enriched ISM. However, the same high pressure should be acting in the opposite direction on the remaining cloud, locally increasing the cloud density, and might even increase the star formation rate at the cloud-blast interface. However, we have not detected a clear evidence of this in the simulation.

The lag being minimum at large galactic radius, as in our example, we expect the SNe to generally explode outside of their dense gaseous nurseries therefore minimizing the role of SN feedback in the destruction of clouds. Less energetic pre-SN effects, active inside the clouds, are not expected to destroy the clumps \emph{per se}, but are likely to modify their inner structure and thus the ongoing star formation. 

This demonstrates the paramount role of the Galactic context (here, the rotation of the disc) as a driver of the evolution of star forming clouds.

\subsection{Clump mass function}
\label{sec:CMF}

\begin{figure}
\includegraphics{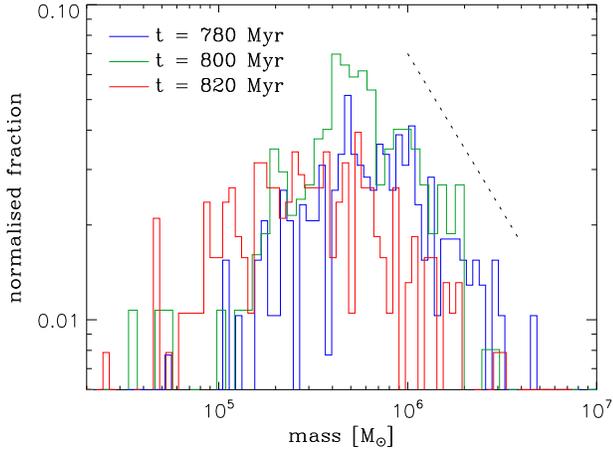}
\caption{Mass function of the gas clouds (denser than $500 \cc$) along their evolution. The black line indicates a power-law of index -1, comparable to the decline of the mass function above $\sim 5\times 10^5 \Msun$. Clumps get less massive with time because of gas depletion due to star formation.}
\label{fig:clump_mf}
\end{figure}

The exploration of the clump mass function (CMF) has become a useful tool to probe the early phase of star formation, and the shape of the IMF. Using CO observations, several authors measured a power-law distribution of the form $\dd N / \dd M \propto M^{-\alpha}$, with an index rather constant across the Galaxy: $\alpha = 1.83$ in the inner disc \citep{Solomon1987}, $\alpha = 1.80$ in the outer Galaxy \citep{Heyer2001} and $\alpha = 1.84$ in high latitude cirrus \citep{Heithausen1998}. (See also \citealt{Reid2010} and \citealt{Hennebelle2012} for a review.) Owing to the diversity of age of the clumps across the Milky Way, this slope is expected to be also constant in time.

In our simulation, the high-mass end of the global CMF shown in \fig{clump_mf} yields a decline of the form $\dd N / \dd \log(M) \propto M^{-\alpha+1}$ (or $\propto M^{-\alpha}$ in linear mass intervals) , with $\alpha \approx 2$, i.e. comparable to the observations. The evolution over $40 \Myr$ does not modify this high-mass behaviour, which implies that the ISM maintains its hierarchical structure \citep{Stutzki1998}, while the maximum of the distribution is shifted towards lighter masses, as already suggested in the beads on a string sample discussed above. We note however that \citet{Audit2010} suggested that the maximum of the CMF might be biased by numerical dissipation.

By arguing that the inner structure of clumps originates from density fluctuations, several authors proposed a relation between the index $\alpha$ of the CMF and that of the power spectrum of the density field  (\citealt{Elmegreen1996, Stutzki1998, Hennebelle2008, Shadmehri2011}, see \citealt{Hennebelle2012} for a review). The hierarchical structure of the ISM in clouds ($10^{3\mh 6} \Msun$), filaments ($10^{2\mh 3} \Msun$) and cores ($10^{0\mh 2} \Msun$) might then be linked to the turbulence cascade, as discussed in the next Section.

\subsection{Power spectrum density and turbulence cascade}
\label{sec:psd}

Together with magnetic fields, which can reduce the star formation efficiency \citep{Federrath2013}, turbulence is one of the key physical contributions to the support of gas structures against gravitation. By knowing at which scale the turbulence is created and how it propagates and diffuses to other scales, one can infer how, where and when the gravity takes over and form structures like discs, clouds, cores or even stars. In simulations like this one, the injection of turbulence at galactic scale is well described, but the resolution remains too low to fully describe the cascade down to the dissipation scale, the so-called Kolmogorov scale \citep{Hennebelle2012}. Our spatial resolution is still too coarse to capture it.

For an \emph{incompressible} fluid, the \citet{Kolmogorov1941} theory predicts an energy spectrum of the velocity field, as a function of the inverse scale-length $k$, of the form $E(k) \propto k^{-5/3}$, hence a power spectrum density (PSD) $P_v(k) \propto k^{-2}E(k) \propto k^{-11/3}$. A pressure-less (i.e. \emph{fully compressible}) fluid would yield a steeper spectrum $P_v(k) \propto k^{-4}$ as proposed by \citet{Burgers1974}, and thus a partially compressible fluid, like the ISM, is expected to lie between the two extremes.

In our case, because of the rotation of the disc, the velocity should be replaced by its dispersion in our spectral analysis. However, the structure of the AMR grid introduces artifacts in the computation of the velocity dispersion where the refinement changes. Futhermore, computing the power spectrum of a three-dimensional field at high resolution and over a large spatial range is involved: for simplicity, we have limited our analysis to a two-dimensional exploration, i.e. a spectrum shallower by an index unity ($\propto k^{-8/3}$ and $k^{-3}$ for the incompressible case and the fully compressible cases, respectively). For a subsonic medium, the power spectrum of the surface density of the gas $\Sigma$ is a good proxy of that of the velocity \citep[see e.g.][]{Burkhart2013}. In the supersonic case however, the conversion is less clear and one should not seek a direct comparison of the spectral indices with that of the Kolmogorov framework. 

\begin{figure}
\includegraphics{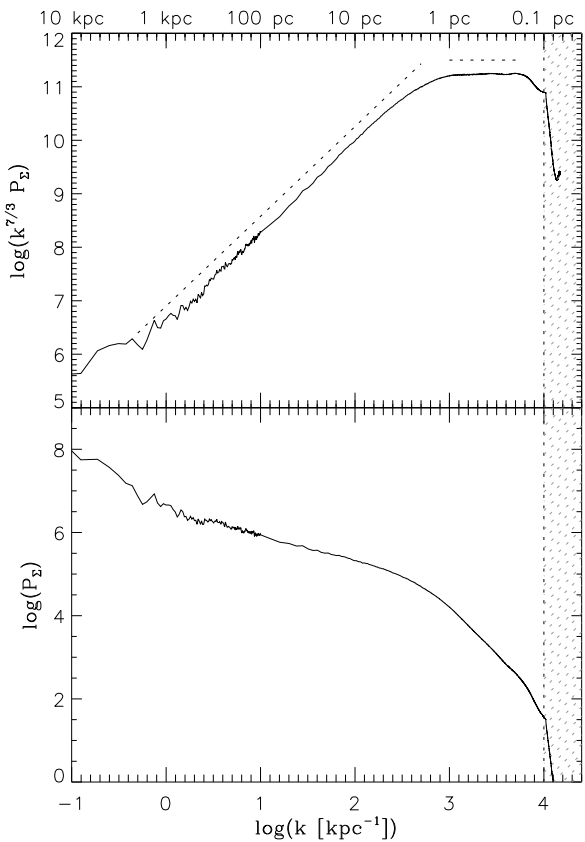}
\caption{Compensated (top) and non-compensated (bottom) power spectrum density of the surface density of the central $8 \kpc \times 8 \kpc$ of the galaxy, constructed from two power spectra at different resolution (and connected at $k=10 \kpc^{-1}$). The two dashed lines represent $P_\Sigma(k) \propto k^{-2/3}$ and $\propto k^{-7/3}$. The shaded area at high $k$'s shows the limit being due to the sampling frequency (two cells = $0.1 \pc$).}
\label{fig:psd}
\end{figure}

\fig{psd} shows the power spectrum density (PSD), computed thanks to a fast Fourier transform of the surface density of gas. The hierarchy of large scales structures ($\gtrsim 2\pc$) sets a power spectrum of index $-0.67 \approx -2/3$ over at least 3 decades. At least at large scale, the ISM of the Milky Way is subsonic and a partially compressible turbulence should yield a PSD of index $< -8/3$. Therefore, finding a shallower PSD indicates the non-turbulent nature of the ISM in this scale range: the gravitation dominates in setting the hierarchy of structures. Note that the scale at which this regime ends cannot be determined by our approach, since the turbulence leaves the subsonic range at an unknown and non-unique scale, voiding the correspondance between the PSD we compute and the Kolmogorov theory.

However, a transition toward a steeper spectrum (index $-2.33 \approx -7/3$) is found at about the parsec scale, i.e. inside the molecular clouds, and over almost a decade. In that case, the ISM is mostly supersonic and the PSD is (presumably) steeper than that from the Kolmogorov regime\footnote{Recall that the $-8/3$ index of the Kolmogorov PSD of the surface density only holds in subsonic gas.}. Thus, once an external gravitational cause (e.g. a spiral shock) has triggered their fragmentation, the clouds develop hydrodynamical turbulence that dominates the PSD, instead of the gravitational hierarchy, as observed at larger scales. Nevertheless, we note that this transition has never been detected in observations nor smaller scale simulations.

When approaching the smallest scale accessible (corresponding to two cells for a spectral analysis, i.e. $0.1 \pc$ in our case), numerical effects appear and bias the power spectrum. The inertial regime of turbulence can only be probed at several times the Nyquist frequency, which in our case, corresponds to a few times\footnote{The exact, unknown, value depends on the numerical implementation and physical properties of the turbulence.} $0.1 \pc$. However, since the transition mentioned above is described by more than 50 resolution elements it is unlikely to be of numerical origin, and it is thus more likely to be a physical effect.

Except from the change of spectral index, no particular features arise from the global PSD. The apparent regularity in the spacing and size of beads on a string (recall Section~\ref{sec:boas}), the width of the spiral arms or the stellar feedback (as \hii bubbles and SN blasts) do not leave any clear signature in the PSD computed over $8\kpc \times 8\kpc$. The diversity of objects and the co-existence of structures at different stages of their evolution (as discussed above) seem to blur the PSD by injecting energy at a great number of scales. In particular, no transition is detected around the scale-height of the disc ($\approx 100 \pc$), contrary to other studies (e.g. the observations by \citealt{Elmegreen2001b} and \citealt{Block2010} or the simulation by \citealt{Bournaud2010b} of the Large Magellanic Cloud, or \citealt{Combes2012} in M33). This difference is discussed below.

\begin{figure}
\includegraphics{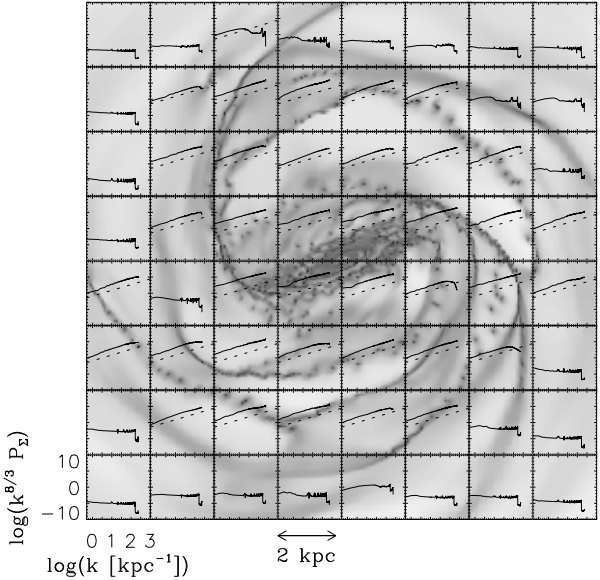}
\caption{Compensated power spectrum density in several regions of the galaxy. Note that the compensation index ($8/3$) is not the same as in \fig{psd} ($7/3$). The dotted lines indicates $P_\Sigma(k) \propto k^{-2/3}$ in the relevant tiles.}
\label{fig:psdmosaic}
\end{figure}

In an attempt to separate un-related structures, \fig{psdmosaic} shows the PSD of the surface density, but computed in $2\kpc \times 2\kpc$ tiles. In the outer part of the galactic disc (e.g. the top-left-most tile), the diffuse gas is isothermal and subsonic (recall \fig{eos} and \ref{fig:pdfmosaic}) and the absence of structures like spirals or clumps allows the turbulence to decay unconstrained. This leads to a Kolmogorov-like power spectra, i.e. $P_\Sigma (k) \propto k^{-8/3}$.

In contrast, the PSDs of the inner regions yield a $-2/3$ index at large scale where the gravitation sets the structure of the ISM, similarly to the global case of \fig{psd}. In some of these regions, the PSD becomes steeper at scales around $1 \mh 10 \pc$, indicating here again that supersonic turbulence dominates. Yet, no injection scale is clearly detected, since each structure influences the PSD at more than one scale. For instance, when a spiral arm crosses a tile, it injects power in the PSD at scales corresponding to its width, but also its length and all scales in between. For beads on a string- or spiral-like structures, this could influence the PSD over more than one decade in wave-numbers. Similarly, the distribution of radii of the \hii bubbles is rather smooth (with a peak at $\approx 0.3 \pc$) and thus covers an extended range of scales, not leading to a clear signature in the PSD, even in a $2\kpc \times 2\kpc$ region.

In many aspects, such PSD differs from that of the Large Magellanic Cloud \citep[LMC,][]{Block2010, Bournaud2010b}. In the Milky Way, the contrast between spirals and interarms is much stronger than in the LMC, mostly because of a higher rotation velocity and thus stronger spiral shocks. Therefore, our model lacks diffuse ISM, which would be responsible for most of the \hi and infrared emission, from an observational point of view. Such a diffuse, pervasive gas would probably be more efficient in revealing characteristic scales, like the thickness of the disc, than the unique dense phase we probe. In their simulation of the LMC, \citet{Bournaud2010b} used the same EOS than the one presented here. As already discussed, such an EOS does not allow for two phases that would favor the existence of a diffuse ISM. Despite this, the very nature of the LMC of having weaker spiral shocks, a less pronounced arm/interarm density contrast than in the Milky Way, and thus the existence of a diffuse ISM allows to probe the thickness of its disc in the PSD, contrarily to our simulation.

\subsection{Regulation of star formation}
\label{sec:sf_regulation}

The structure and physical properties of the ISM described above lead to a global SFR of $\sim 1 \mh 5 \Msun\yr^{-1}$ measured in the simulation. In absence of magnetic fields, turbulence support and feedback are the main physical regulators of star formation. Furthermore, in this simulation, the artificial fragmentation of the gas due to our finite resolution is compensated by introducing a thermal pressure floor (the Jeans polytrope, recall Section~\ref{sec:eos}). This necessary artifact could also regulate star formation. However, the maximum heating we measure is moderate ($1600 \U{K}$) and only one percent of the gas mass is heated above $100 \U{K}$. At this temperature, the sound speed ($\sim 1 \kms$)\footnote{We note that the thermal velocity dispersion introduced by the polytrope are in agreement with the non-thermal values measured in molecular clouds, at the parsec and subparsec scales (see \citealt{Hennebelle2012}, their figure 5).} remains well below the turbulent velocities. Therefore, the regulation of star formation by the polytrope concerns only a small fraction of the gas, and is negligible with respect to supersonic turbulence.

As we have seen in Section~\ref{sec:drift}, at least in some cases, most of the stellar feedback energy is released outside of the clouds, and thus plays a secondary role in regulating star formation in these volumes. Supersonic turbulence should therefore be the main support against rapid collapse of the clouds and further star formation. For example, the velocity dispersion of the clouds in spurs of a few $10^5 \Msun$ is $\sim 10 \mh 40 \kms$, i.e. high enough to prevent most of the collapse into dense cores.

In other regions, like at tips of the bar where the number density of cloud and the inter-cloud density are several order of magnitude higher, the role of feedback might be more important (but more difficult to probe owing to the rapid variations of the density field), and adds to the turbulent support.

Finally, \emph{inside} the bar (recall Section~\ref{sec:ic}), phenomena external to the cloud itself like shear and galactic tides govern the organisation of the ISM by preventing the very collapse of gas around dense seeds, which significantly reduces the star formation activity in this area.

Although supersonic turbulence often dominates, secondary aspects like feedback, shear and tides also participate in regulating star formation. Their relative weights in this process depend on the environment of the cloud, and varies significantly across the galaxy, as noted from an observational point of view by \citet{Meidt2013}. We suspect that such variations would be amplified in more extreme environments like colliding galaxies.

\section{Summary and conclusions}

We presented a simulation of a grand-design spiral galaxy with properties comparable to those of the Milky Way. Although some structures like beads on a string seem over-represented compared to the real Galaxy, we argue that the match is likely to improve at a later stage of the evolution, not reached here. The major interest of this work is the possibility of probing physical quantities and structures at a sub-parsec resolution, in a fully consistent galactic context. This resolution allows us to avoid many artifacts and biases introduced in previous studies at lower resolution. Many topics and questions will be adressed in forthcoming contributions using this simulation, but for now, the main results presented here are as follows:
\begin{itemize}
\item The stellar and gaseous spirals extend further than the outer Lindblad resonance ($6.3 \kpc$) as determined from the pattern speed of the bar. This resonance seems to only affect their pitch angle. However, a secondary pattern speed associated with the spirals themselves explains their truncation at $8 \kpc$.
\item The formation of a bar triggers the appearance of inner Lindblad resonances (40 and $450 \pc$) which lead to the accumulation of gas at these radii. A nuclear disc forms around the SMBH and controls its fueling.
\item The gas density PDF can be approximated to first order by a log-normal functional form with a power-law tail at high density ($ \gtrsim 2000 \cc$), where the gaseous clumps are self-gravitating. The index of the power-law is connected to the radial density profile of the clouds, as observed.
\item Describing the transition from turbulence supported to self-gravitating gas is a necessary condition for resolving star formation. Such transition would ideally be used as a density threshold in numerical implementations of star formation.
\item The gas in some spiral arms is organised as beads on a string, with a relatively regular spacing. Because mostly of star formation the mass and the size of these clumps decrease with time, while their separation increases.
\item Along the most pronounced gas spirals, a velocity gradient across the arm generates Kelvin-Helmholtz instabilities in the form of spurs hosting star formation in the convex (leading) side of the long, continuous, ``grand-design'' arms. The thin gas tail connecting the cloud to its spiral arm dissolves in $\sim 15 \Myr$, while the dense cloud survives.
\item The signatures of stellar feedback in the density field are clearly visible in the vicinity of the beads. The asymmetric drift linked to galactic rotation is responsible for an offset between the locii of the SN explosions and the densest regions of their clouds. Therefore, because of large scale effects, the stellar feedback is not efficient at destroying such clouds.
\item The high-mass end of the clump mass function declines as a unity-index power-law function of the mass. This slope remains unchanged as the average mass and radius decrease in time. This may evolve into the star cluster mass function.
\item The regulation of star formation at $\lesssim 5 \Msun\yr^{-1}$, in spite of a star formation efficiency of 3 percent, is ensured by the supersonic turbulent support over a broad range of densities, and only in a much milder way by stellar feedback.
\end{itemize}

The computational cost of the resolution of this simulation limits the duration of the simulation: our analysis at sub-parsec resolution spans only a few $\sim 10 \Myr$. The evolution of large scale structures such as the bar and spiral arms, and global phenomena such as outflows, cannot be followed here. However, thanks to a realistic model at galactic scale, we are able to determine the behaviour of gaseous features of smaller sizes and the conditions for star formation. Future contributions will focus on several phenomena and structures accessible in this simulation: e.g. the laws of star formation, the physics of the galactic nucleus and the inner structures of clouds.

\section*{Acknowledgments}
We thank S\'ebastien Fromang, Steve Longmore and Pierre-Alain Duc for interesting discussions, and the anonymous referee for providing an insightful report. This work was granted access to the PRACE Research Infrastructure resource \emph{Curie} hosted at the TGCC (France). FR, FB and JG acknowledge support from the EC through grant ERC-StG-257720. AD acknowledges support from ISF grant 24/12, GIF grant G-1052-104.7/2009, a DIP-DFG grant, and NSF grant AST-1010033.

\bibliographystyle{mn2e}

\begin{thebibliography}{}

\bibitem[\protect\citeauthoryear{{Aarseth}}{{Aarseth}}{2003}]{Aarseth2003}
{Aarseth} S.~J.,  2003, {Gravitational N-Body Simulations}.
{Cambridge University Press, November 2003.}

\bibitem[\protect\citeauthoryear{{Arzoumanian}, {Andr{\'e}}, {Didelon},
  {K{\"o}nyves}, {Schneider}, {Men'shchikov}, {Sousbie}, {Zavagno} \& {et
  al.}}{{Arzoumanian} et~al.}{2011}]{Arzoumanian2011}
{Arzoumanian} D.,  {Andr{\'e}} P.,  {Didelon} P.,  {K{\"o}nyves} V.,
  {Schneider} N.,  {Men'shchikov} A.,  {Sousbie} T.,  {Zavagno}   {et al.}
  2011, \aap, 529, L6

\bibitem[\protect\citeauthoryear{{Aubert} \& {Teyssier}}{{Aubert} \&
  {Teyssier}}{2008}]{Aubert2008}
{Aubert} D.,  {Teyssier} R.,  2008, \mnras, 387, 295

\bibitem[\protect\citeauthoryear{{Audit} \& {Hennebelle}}{{Audit} \&
  {Hennebelle}}{2010}]{Audit2010}
{Audit} E.,  {Hennebelle} P.,  2010, \aap, 511, A76

\bibitem[\protect\citeauthoryear{{Barnes}, {van Zee}, {C{\^o}t{\'e}} \&
  {Schade}}{{Barnes} et~al.}{2012}]{Barnes2012}
{Barnes} K.~L.,  {van Zee} L.,  {C{\^o}t{\'e}} S.,    {Schade} D.,  2012, \apj,
  757, 64

\bibitem[\protect\citeauthoryear{{Bate} \& {Bonnell}}{{Bate} \&
  {Bonnell}}{2005}]{Bate2005}
{Bate} M.~R.,  {Bonnell} I.~A.,  2005, \mnras, 356, 1201

\bibitem[\protect\citeauthoryear{{Bate}, {Bonnell} \& {Price}}{{Bate}
  et~al.}{1995}]{Bate1995}
{Bate} M.~R.,  {Bonnell} I.~A.,    {Price} N.~M.,  1995, \mnras, 277, 362

\bibitem[\protect\citeauthoryear{{Bekki}}{{Bekki}}{2012}]{Bekki2012}
{Bekki} K.,  2012, \mnras, 422, 1957

\bibitem[\protect\citeauthoryear{{Bendinelli}}{{Bendinelli}}{1991}]{Bendinelli1991}
{Bendinelli} O.,  1991, \apj, 366, 599

\bibitem[\protect\citeauthoryear{{Binney} \& {Tremaine}}{{Binney} \&
  {Tremaine}}{2008}]{Binney2008}
{Binney} J.,  {Tremaine} S.,  2008, {Galactic Dynamics: Second Edition}.
Princeton University Press

\bibitem[\protect\citeauthoryear{{Block}, {Puerari}, {Elmegreen} \&
  {Bournaud}}{{Block} et~al.}{2010}]{Block2010}
{Block} D.~L.,  {Puerari} I.,  {Elmegreen} B.~G.,    {Bournaud} F.,  2010,
  \apjl, 718, L1

\bibitem[\protect\citeauthoryear{{Bonnell}, {Dobbs}, {Robitaille} \&
  {Pringle}}{{Bonnell} et~al.}{2006}]{Bonnell2006}
{Bonnell} I.~A.,  {Dobbs} C.~L.,  {Robitaille} T.~P.,    {Pringle} J.~E.,
  2006, \mnras, 365, 37

\bibitem[\protect\citeauthoryear{{Bonnell}, {Dobbs} \& {Smith}}{{Bonnell}
  et~al.}{2013}]{Bonnell2013}
{Bonnell} I.~A.,  {Dobbs} C.~L.,    {Smith} R.~J.,  2013, \mnras, 430, 1790

\bibitem[\protect\citeauthoryear{{Bournaud}, {Elmegreen}, {Teyssier}, {Block}
  \& {Puerari}}{{Bournaud} et~al.}{2010}]{Bournaud2010b}
{Bournaud} F.,  {Elmegreen} B.~G.,  {Teyssier} R.,  {Block} D.~L.,    {Puerari}
  I.,  2010, \mnras, 409, 1088

\bibitem[\protect\citeauthoryear{Burgers}{Burgers}{1974}]{Burgers1974}
Burgers J.,  1974, The Non-Linear Diffusion Equation: Asymptotic Solutions and
  Statistical Problems.
Lecture series, Springer

\bibitem[\protect\citeauthoryear{{Burkhart}, {Lazarian}, {Ossenkopf} \&
  {Stutzki}}{{Burkhart} et~al.}{2013}]{Burkhart2013}
{Burkhart} B.,  {Lazarian} A.,  {Ossenkopf} V.,    {Stutzki} J.,  2013, \apj,
  771, 123

\bibitem[\protect\citeauthoryear{{Chakrabarti}, {Laughlin} \&
  {Shu}}{{Chakrabarti} et~al.}{2003}]{Chakrabarti2003}
{Chakrabarti} S.,  {Laughlin} G.,    {Shu} F.~H.,  2003, \apj, 596, 220

\bibitem[\protect\citeauthoryear{{Chevalier}}{{Chevalier}}{1989}]{Chevalier1989}
{Chevalier} R.~A.,  1989, \apj, 346, 847

\bibitem[\protect\citeauthoryear{{Combes}, {Boquien}, {Kramer}, {Xilouris},
  {Bertoldi}, {Braine}, {Buchbender}, {Calzetti} \& {et al.}}{{Combes}
  et~al.}{2012}]{Combes2012}
{Combes} F.,  {Boquien} M.,  {Kramer} C.,  {Xilouris} E.~M.,  {Bertoldi} F.,
  {Braine} J.,  {Buchbender} C.,  {Calzetti} D.,    {et al.} 2012, \aap, 539,
  A67

\bibitem[\protect\citeauthoryear{{Courant}, {Friedrichs} \& {Lewy}}{{Courant}
  et~al.}{1928}]{Courant1928}
{Courant} R.,  {Friedrichs} K.,    {Lewy} H.,  1928, Mathematische Annalen,
  100, 32

\bibitem[\protect\citeauthoryear{{Dale} \& {Bonnell}}{{Dale} \&
  {Bonnell}}{2011}]{Dale2011}
{Dale} J.~E.,  {Bonnell} I.,  2011, \mnras, 414, 321

\bibitem[\protect\citeauthoryear{{Dale}, {Bonnell}, {Clarke} \& {Bate}}{{Dale}
  et~al.}{2005}]{Dale2005}
{Dale} J.~E.,  {Bonnell} I.~A.,  {Clarke} C.~J.,    {Bate} M.~R.,  2005,
  \mnras, 358, 291

\bibitem[\protect\citeauthoryear{{Dame}, {Hartmann} \& {Thaddeus}}{{Dame}
  et~al.}{2001}]{Dame2001}
{Dame} T.~M.,  {Hartmann} D.,    {Thaddeus} P.,  2001, \apj, 547, 792

\bibitem[\protect\citeauthoryear{{Debattista}, {Gerhard} \&
  {Sevenster}}{{Debattista} et~al.}{2002}]{Debattista2002}
{Debattista} V.~P.,  {Gerhard} O.,    {Sevenster} M.~N.,  2002, \mnras, 334,
  355

\bibitem[\protect\citeauthoryear{{Dekel} \& {Birnboim}}{{Dekel} \&
  {Birnboim}}{2006}]{Dekel2006}
{Dekel} A.,  {Birnboim} Y.,  2006, \mnras, 368, 2

\bibitem[\protect\citeauthoryear{{Dekel} \& {Krumholz}}{{Dekel} \&
  {Krumholz}}{2013}]{Dekel2013}
{Dekel} A.,  {Krumholz} M.~R.,  2013, \mnras, 432, 455

\bibitem[\protect\citeauthoryear{{Di Matteo}, {Combes}, {Melchior} \&
  {Semelin}}{{Di Matteo} et~al.}{2007}]{diMatteo2007}
{Di Matteo} P.,  {Combes} F.,  {Melchior} A.,    {Semelin} B.,  2007, \aap,
  468, 61

\bibitem[\protect\citeauthoryear{{Dobbs}}{{Dobbs}}{2008}]{Dobbs2008}
{Dobbs} C.~L.,  2008, \mnras, 391, 844

\bibitem[\protect\citeauthoryear{{Dobbs} \& {Bonnell}}{{Dobbs} \&
  {Bonnell}}{2006}]{Dobbs2006}
{Dobbs} C.~L.,  {Bonnell} I.~A.,  2006, \mnras, 367, 873

\bibitem[\protect\citeauthoryear{{Dobbs}, {Pringle} \& {Burkert}}{{Dobbs}
  et~al.}{2012}]{Dobbs2012}
{Dobbs} C.~L.,  {Pringle} J.~E.,    {Burkert} A.,  2012, \mnras, 425, 2157

\bibitem[\protect\citeauthoryear{{Dubois} \& {Teyssier}}{{Dubois} \&
  {Teyssier}}{2008}]{Dubois2008}
{Dubois} Y.,  {Teyssier} R.,  2008, \aap, 477, 79

\bibitem[\protect\citeauthoryear{{Elmegreen}}{{Elmegreen}}{2011}]{Elmegreen2011}
{Elmegreen} B.~G.,  2011, \apj, 731, 61

\bibitem[\protect\citeauthoryear{{Elmegreen} \& {Elmegreen}}{{Elmegreen} \&
  {Elmegreen}}{1983}]{Elmegreen1983}
{Elmegreen} B.~G.,  {Elmegreen} D.~M.,  1983, \mnras, 203, 31

\bibitem[\protect\citeauthoryear{{Elmegreen} \& {Falgarone}}{{Elmegreen} \&
  {Falgarone}}{1996}]{Elmegreen1996}
{Elmegreen} B.~G.,  {Falgarone} E.,  1996, \apj, 471, 816

\bibitem[\protect\citeauthoryear{{Elmegreen}, {Kim} \&
  {Staveley-Smith}}{{Elmegreen} et~al.}{2001}]{Elmegreen2001b}
{Elmegreen} B.~G.,  {Kim} S.,    {Staveley-Smith} L.,  2001, \apj, 548, 749

\bibitem[\protect\citeauthoryear{{Elmegreen}}{{Elmegreen}}{1980}]{Elmegreen1980}
{Elmegreen} D.~M.,  1980, \apj, 242, 528

\bibitem[\protect\citeauthoryear{{Emsellem}, {Monnet} \& {Bacon}}{{Emsellem}
  et~al.}{1994}]{Emsellem1994}
{Emsellem} E.,  {Monnet} G.,    {Bacon} R.,  1994, \aap, 285, 723

\bibitem[\protect\citeauthoryear{{Federrath} \& {Klessen}}{{Federrath} \&
  {Klessen}}{2012}]{Federrath2012}
{Federrath} C.,  {Klessen} R.~S.,  2012, \apj, 761, 156

\bibitem[\protect\citeauthoryear{{Federrath} \& {Klessen}}{{Federrath} \&
  {Klessen}}{2013}]{Federrath2013}
{Federrath} C.,  {Klessen} R.~S.,  2013, \apj, 763, 51

\bibitem[\protect\citeauthoryear{{Federrath}, {Klessen} \&
  {Schmidt}}{{Federrath} et~al.}{2008}]{Federrath2008}
{Federrath} C.,  {Klessen} R.~S.,    {Schmidt} W.,  2008, \apjl, 688, L79

\bibitem[\protect\citeauthoryear{{Federrath}, {Roman-Duval}, {Klessen},
  {Schmidt} \& {Mac Low}}{{Federrath} et~al.}{2010}]{Federrath2010}
{Federrath} C.,  {Roman-Duval} J.,  {Klessen} R.~S.,  {Schmidt} W.,    {Mac
  Low} M.-M.,  2010, \aap, 512, A81

\bibitem[\protect\citeauthoryear{{Fischera} \& {Martin}}{{Fischera} \&
  {Martin}}{2012}]{Fischera2012}
{Fischera} J.,  {Martin} P.~G.,  2012, \aap, 542, A77

\bibitem[\protect\citeauthoryear{{Guedes}, {Callegari}, {Madau} \&
  {Mayer}}{{Guedes} et~al.}{2011}]{Guedes2011}
{Guedes} J.,  {Callegari} S.,  {Madau} P.,    {Mayer} L.,  2011, \apj, 742, 76

\bibitem[\protect\citeauthoryear{{Guesten} \& {Mezger}}{{Guesten} \&
  {Mezger}}{1982}]{Guesten1982}
{Guesten} R.,  {Mezger} P.~G.,  1982, Vistas in Astronomy, 26, 159

\bibitem[\protect\citeauthoryear{{Haardt} \& {Madau}}{{Haardt} \&
  {Madau}}{1996}]{Haardt1996}
{Haardt} F.,  {Madau} P.,  1996, \apj, 461, 20

\bibitem[\protect\citeauthoryear{{Heithausen}, {Bensch}, {Stutzki}, {Falgarone}
  \& {Panis}}{{Heithausen} et~al.}{1998}]{Heithausen1998}
{Heithausen} A.,  {Bensch} F.,  {Stutzki} J.,  {Falgarone} E.,    {Panis}
  J.~F.,  1998, \aap, 331, L65

\bibitem[\protect\citeauthoryear{{Hennebelle} \& {Chabrier}}{{Hennebelle} \&
  {Chabrier}}{2008}]{Hennebelle2008}
{Hennebelle} P.,  {Chabrier} G.,  2008, \apj, 684, 395

\bibitem[\protect\citeauthoryear{{Hennebelle} \& {Falgarone}}{{Hennebelle} \&
  {Falgarone}}{2012}]{Hennebelle2012}
{Hennebelle} P.,  {Falgarone} E.,  2012, \aapr, 20, 55

\bibitem[\protect\citeauthoryear{{Heyer}, {Carpenter} \& {Snell}}{{Heyer}
  et~al.}{2001}]{Heyer2001}
{Heyer} M.~H.,  {Carpenter} J.~M.,    {Snell} R.~L.,  2001, \apj, 551, 852

\bibitem[\protect\citeauthoryear{{Hopkins}}{{Hopkins}}{2013}]{Hopkins2013}
{Hopkins} P.~F.,  2013, \mnras, 430, 1880

\bibitem[\protect\citeauthoryear{{Hopkins}, {Quataert} \& {Murray}}{{Hopkins}
  et~al.}{2011}]{Hopkins2011}
{Hopkins} P.~F.,  {Quataert} E.,    {Murray} N.,  2011, \mnras, 417, 950

\bibitem[\protect\citeauthoryear{{Hopkins}, {Quataert} \& {Murray}}{{Hopkins}
  et~al.}{2012}]{Hopkins2012}
{Hopkins} P.~F.,  {Quataert} E.,    {Murray} N.,  2012, \mnras, 421, 3488

\bibitem[\protect\citeauthoryear{{Kainulainen}, {Federrath} \&
  {Henning}}{{Kainulainen} et~al.}{2013}]{Kainulainen2013}
{Kainulainen} J.,  {Federrath} C.,    {Henning} T.,  2013, \aap, 553, L8

\bibitem[\protect\citeauthoryear{{Karl}, {Naab}, {Johansson}, {Kotarba},
  {Boily}, {Renaud} \& {Theis}}{{Karl} et~al.}{2010}]{Karl2010}
{Karl} S.~J.,  {Naab} T.,  {Johansson} P.~H.,  {Kotarba} H.,  {Boily} C.~M.,
  {Renaud} F.,    {Theis} C.,  2010, \apjl, 715, L88

\bibitem[\protect\citeauthoryear{{Katz}}{{Katz}}{1992}]{Katz1992}
{Katz} N.,  1992, \apj, 391, 502

\bibitem[\protect\citeauthoryear{{Khoperskov}, {Vasiliev}, {Sobolev} \&
  {Khoperskov}}{{Khoperskov} et~al.}{2013}]{Khoperskov2013}
{Khoperskov} S.~A.,  {Vasiliev} E.~O.,  {Sobolev} A.~M.,    {Khoperskov} A.~V.,
   2013, \mnras, 428, 2311

\bibitem[\protect\citeauthoryear{{Kim} \& {Ostriker}}{{Kim} \&
  {Ostriker}}{2006}]{Kim2006}
{Kim} W.-T.,  {Ostriker} E.~C.,  2006, \apj, 646, 213

\bibitem[\protect\citeauthoryear{{Kim}, {Ostriker} \& {Stone}}{{Kim}
  et~al.}{2002}]{Kim2002}
{Kim} W.-T.,  {Ostriker} E.~C.,    {Stone} J.~M.,  2002, \apj, 581, 1080

\bibitem[\protect\citeauthoryear{{Klessen}}{{Klessen}}{2000}]{Klessen2000}
{Klessen} R.~S.,  2000, \apj, 535, 869

\bibitem[\protect\citeauthoryear{{Kolmogorov}}{{Kolmogorov}}{1941}]{Kolmogorov1941}
{Kolmogorov} A.,  1941, Akademiia Nauk SSSR Doklady, 30, 301

\bibitem[\protect\citeauthoryear{{Kraljic}, {Bournaud} \& {Martig}}{{Kraljic}
  et~al.}{2012}]{Kraljic2012}
{Kraljic} K.,  {Bournaud} F.,    {Martig} M.,  2012, \apj, 757, 60

\bibitem[\protect\citeauthoryear{{Krumholz} \& {Dekel}}{{Krumholz} \&
  {Dekel}}{2010}]{Krumholz2010}
{Krumholz} M.~R.,  {Dekel} A.,  2010, \mnras, 406, 112

\bibitem[\protect\citeauthoryear{{Krumholz}, {Dekel} \& {McKee}}{{Krumholz}
  et~al.}{2012}]{Krumholz2012}
{Krumholz} M.~R.,  {Dekel} A.,    {McKee} C.~F.,  2012, \apj, 745, 69

\bibitem[\protect\citeauthoryear{{Krumholz}, {McKee} \& {Tumlinson}}{{Krumholz}
  et~al.}{2009}]{Krumholz2009a}
{Krumholz} M.~R.,  {McKee} C.~F.,    {Tumlinson} J.,  2009, \apj, 699, 850

\bibitem[\protect\citeauthoryear{{Krumholz} \& {Tan}}{{Krumholz} \&
  {Tan}}{2007}]{Krumholz2007a}
{Krumholz} M.~R.,  {Tan} J.~C.,  2007, \apj, 654, 304

\bibitem[\protect\citeauthoryear{{La Vigne}, {Vogel} \& {Ostriker}}{{La Vigne}
  et~al.}{2006}]{LaVigne2006}
{La Vigne} M.~A.,  {Vogel} S.~N.,    {Ostriker} E.~C.,  2006, \apj, 650, 818

\bibitem[\protect\citeauthoryear{{Lombardi}, {Lada} \& {Alves}}{{Lombardi}
  et~al.}{2008}]{Lombardi2008}
{Lombardi} M.,  {Lada} C.~J.,    {Alves} J.,  2008, \aap, 489, 143

\bibitem[\protect\citeauthoryear{{Lombardi}, {Lada} \& {Alves}}{{Lombardi}
  et~al.}{2010}]{Lombardi2010}
{Lombardi} M.,  {Lada} C.~J.,    {Alves} J.,  2010, \aap, 512, A67

\bibitem[\protect\citeauthoryear{{Meidt}, {Schinnerer}, {Garcia-Burillo},
  {Hughes}, {Colombo}, {Pety}, {Dobbs}, {Schuster}, {Kramer}, {Leroy}, {Dumas}
  \& {Thompson}}{{Meidt} et~al.}{2013}]{Meidt2013}
{Meidt} S.~E.,  {Schinnerer} E.,  {Garcia-Burillo} S.,  {Hughes} A.,  {Colombo}
  D.,  {Pety} J.,  {Dobbs} C.~L.,  {Schuster} K.~F.,  {Kramer} C.,  {Leroy}
  A.~K.,  {Dumas} G.,    {Thompson} T.~A.,  2013, ArXiv e-prints

\bibitem[\protect\citeauthoryear{{Molina}, {Glover}, {Federrath} \&
  {Klessen}}{{Molina} et~al.}{2012}]{Molina2012}
{Molina} F.~Z.,  {Glover} S.~C.~O.,  {Federrath} C.,    {Klessen} R.~S.,  2012,
  \mnras, 423, 2680

\bibitem[\protect\citeauthoryear{{Monnet}, {Bacon} \& {Emsellem}}{{Monnet}
  et~al.}{1992}]{Monnet1992}
{Monnet} G.,  {Bacon} R.,    {Emsellem} E.,  1992, \aap, 253, 366

\bibitem[\protect\citeauthoryear{{Nordlund} \& {Padoan}}{{Nordlund} \&
  {Padoan}}{1999}]{Nordlund1999}
{Nordlund} {\AA}.~K.,  {Padoan} P.,  1999, in {Franco} J.,  {Carraminana} A.,
  eds, Interstellar Turbulence {The Density PDFs of Supersonic Random Flows}.
p.~218

\bibitem[\protect\citeauthoryear{{Norman} \& {Silk}}{{Norman} \&
  {Silk}}{1980}]{Norman1980}
{Norman} C.,  {Silk} J.,  1980, \apj, 238, 158

\bibitem[\protect\citeauthoryear{{Oppenheimer} \& {Dav{\'e}}}{{Oppenheimer} \&
  {Dav{\'e}}}{2006}]{Oppenheimer2006}
{Oppenheimer} B.~D.,  {Dav{\'e}} R.,  2006, \mnras, 373, 1265

\bibitem[\protect\citeauthoryear{{Oppenheimer}, {Dav{\'e}}, {Kere{\v s}},
  {Fardal}, {Katz}, {Kollmeier} \& {Weinberg}}{{Oppenheimer}
  et~al.}{2010}]{Oppenheimer2010}
{Oppenheimer} B.~D.,  {Dav{\'e}} R.,  {Kere{\v s}} D.,  {Fardal} M.,  {Katz}
  N.,  {Kollmeier} J.~A.,    {Weinberg} D.~H.,  2010, \mnras, 406, 2325

\bibitem[\protect\citeauthoryear{{Padoan}, {Juvela}, {Goodman} \&
  {Nordlund}}{{Padoan} et~al.}{2001}]{Padoan2001}
{Padoan} P.,  {Juvela} M.,  {Goodman} A.~A.,    {Nordlund} {\AA}.,  2001, \apj,
  553, 227

\bibitem[\protect\citeauthoryear{{Padoan}, {Juvela}, {Kritsuk} \&
  {Norman}}{{Padoan} et~al.}{2009}]{Padoan2009}
{Padoan} P.,  {Juvela} M.,  {Kritsuk} A.,    {Norman} M.~L.,  2009, \apjl, 707,
  L153

\bibitem[\protect\citeauthoryear{{Padoan} \& {Nordlund}}{{Padoan} \&
  {Nordlund}}{2011}]{Padoan2011}
{Padoan} P.,  {Nordlund} {\AA}.,  2011, \apj, 730, 40

\bibitem[\protect\citeauthoryear{{Pakmor} \& {Springel}}{{Pakmor} \&
  {Springel}}{2013}]{Pakmor2013}
{Pakmor} R.,  {Springel} V.,  2013, \mnras, 432, 176

\bibitem[\protect\citeauthoryear{{Povich}}{{Povich}}{2012}]{Povich2012}
{Povich} M.~S.,  2012, ArXiv e-prints

\bibitem[\protect\citeauthoryear{{Reid}, {Wadsley}, {Petitclerc} \&
  {Sills}}{{Reid} et~al.}{2010}]{Reid2010}
{Reid} M.~A.,  {Wadsley} J.,  {Petitclerc} N.,    {Sills} A.,  2010, \apj, 719,
  561

\bibitem[\protect\citeauthoryear{{Renaud}, {Gieles} \& {Boily}}{{Renaud}
  et~al.}{2011}]{Renaud2011}
{Renaud} F.,  {Gieles} M.,    {Boily} C.~M.,  2011, \mnras, 418, 759

\bibitem[\protect\citeauthoryear{{Renaud}, {Kraljic} \& {Bournaud}}{{Renaud}
  et~al.}{2012}]{Renaud2012}
{Renaud} F.,  {Kraljic} K.,    {Bournaud} F.,  2012, \apjl, 760, L16

\bibitem[\protect\citeauthoryear{{Robertson} \& {Kravtsov}}{{Robertson} \&
  {Kravtsov}}{2008}]{Robertson2008}
{Robertson} B.~E.,  {Kravtsov} A.~V.,  2008, \apj, 680, 1083

\bibitem[\protect\citeauthoryear{{Robin}, {Reyl{\'e}}, {Derri{\`e}re} \&
  {Picaud}}{{Robin} et~al.}{2003}]{Robin2003}
{Robin} A.~C.,  {Reyl{\'e}} C.,  {Derri{\`e}re} S.,    {Picaud} S.,  2003,
  \aap, 409, 523

\bibitem[\protect\citeauthoryear{{Robitaille} \& {Whitney}}{{Robitaille} \&
  {Whitney}}{2010}]{Robitaille2010}
{Robitaille} T.~P.,  {Whitney} B.~A.,  2010, \apjl, 710, L11

\bibitem[\protect\citeauthoryear{{Salpeter}}{{Salpeter}}{1955}]{Salpeter1955}
{Salpeter} E.~E.,  1955, \apj, 121, 161

\bibitem[\protect\citeauthoryear{{Schaye}}{{Schaye}}{2004}]{Schaye2004}
{Schaye} J.,  2004, \apj, 609, 667

\bibitem[\protect\citeauthoryear{{Schinnerer}, {Meidt}, {Pety}, {Hughes},
  {Colombo}, {Garcia-Burillo}, {Schuster}, {Dumas} \& {et al.}}{{Schinnerer}
  et~al.}{2013}]{Schinnerer2013}
{Schinnerer} E.,  {Meidt} S.~E.,  {Pety} J.,  {Hughes} A.,  {Colombo} D.,
  {Garcia-Burillo} S.,  {Schuster} K.~F.,  {Dumas} G.,    {et al.} 2013, ArXiv
  e-prints

\bibitem[\protect\citeauthoryear{{Schmidt}}{{Schmidt}}{1959}]{Schmidt1959}
{Schmidt} M.,  1959, \apj, 129, 243

\bibitem[\protect\citeauthoryear{{Sellwood} \& {Sparke}}{{Sellwood} \&
  {Sparke}}{1988}]{SellWood1988}
{Sellwood} J.~A.,  {Sparke} L.~S.,  1988, \mnras, 231, 25P

\bibitem[\protect\citeauthoryear{{Shadmehri} \& {Elmegreen}}{{Shadmehri} \&
  {Elmegreen}}{2011}]{Shadmehri2011}
{Shadmehri} M.,  {Elmegreen} B.~G.,  2011, \mnras, 410, 788

\bibitem[\protect\citeauthoryear{{Shetty} \& {Ostriker}}{{Shetty} \&
  {Ostriker}}{2006}]{Shetty2006}
{Shetty} R.,  {Ostriker} E.~C.,  2006, \apj, 647, 997

\bibitem[\protect\citeauthoryear{{Shu}}{{Shu}}{1977}]{Shu1977}
{Shu} F.~H.,  1977, \apj, 214, 488

\bibitem[\protect\citeauthoryear{{Shu}, {Adams} \& {Lizano}}{{Shu}
  et~al.}{1987}]{Shu1987}
{Shu} F.~H.,  {Adams} F.~C.,    {Lizano} S.,  1987, \araa, 25, 23

\bibitem[\protect\citeauthoryear{{Silva-Villa} \& {Larsen}}{{Silva-Villa} \&
  {Larsen}}{2012}]{SilvaVilla2012}
{Silva-Villa} E.,  {Larsen} S.~S.,  2012, \aap, 537, A145

\bibitem[\protect\citeauthoryear{{Solomon}, {Rivolo}, {Barrett} \&
  {Yahil}}{{Solomon} et~al.}{1987}]{Solomon1987}
{Solomon} P.~M.,  {Rivolo} A.~R.,  {Barrett} J.,    {Yahil} A.,  1987, \apj,
  319, 730

\bibitem[\protect\citeauthoryear{{Str{\"o}mgren}}{{Str{\"o}mgren}}{1939}]{Stroemgren1939}
{Str{\"o}mgren} B.,  1939, \apj, 89, 526

\bibitem[\protect\citeauthoryear{{Stutzki}, {Bensch}, {Heithausen}, {Ossenkopf}
  \& {Zielinsky}}{{Stutzki} et~al.}{1998}]{Stutzki1998}
{Stutzki} J.,  {Bensch} F.,  {Heithausen} A.,  {Ossenkopf} V.,    {Zielinsky}
  M.,  1998, \aap, 336, 697

\bibitem[\protect\citeauthoryear{{Tasker}}{{Tasker}}{2011}]{Tasker2011}
{Tasker} E.~J.,  2011, \apj, 730, 11

\bibitem[\protect\citeauthoryear{{Teyssier}}{{Teyssier}}{2002}]{Teyssier2002}
{Teyssier} R.,  2002, \aap, 385, 337

\bibitem[\protect\citeauthoryear{{Teyssier}, {Chapon} \& {Bournaud}}{{Teyssier}
  et~al.}{2010}]{Teyssier2010}
{Teyssier} R.,  {Chapon} D.,    {Bournaud} F.,  2010, \apjl, 720, L149

\bibitem[\protect\citeauthoryear{{Thilker}, {Bianchi}, {Meurer}, {Gil de Paz},
  {Boissier}, {Madore}, {Boselli}, {Ferguson} \& {et al.}}{{Thilker}
  et~al.}{2007}]{Thilker2007}
{Thilker} D.~A.,  {Bianchi} L.,  {Meurer} G.,  {Gil de Paz} A.,  {Boissier} S.,
   {Madore} B.~F.,  {Boselli} A.,  {Ferguson} A.~M.~N.,    {et al.} 2007,
  \apjs, 173, 538

\bibitem[\protect\citeauthoryear{{Thomas}, {Zaroubi}, {Ciardi}, {Pawlik},
  {Labropoulos}, {Jeli{\'c}}, {Bernardi}, {Brentjens}, {de Bruyn}, {Harker},
  {Koopmans}, {Mellema}, {Pandey}, {Schaye} \& {Yatawatta}}{{Thomas}
  et~al.}{2009}]{Thomas2009}
{Thomas} R.~M.,  {Zaroubi} S.,  {Ciardi} B.,  {Pawlik} A.~H.,  {Labropoulos}
  P.,  {Jeli{\'c}} V.,  {Bernardi} G.,  {Brentjens} M.~A.,  {de Bruyn} A.~G.,
  {Harker} G.~J.~A.,  {Koopmans} L.~V.~E.,  {Mellema} G.,  {Pandey} V.~N.,
  {Schaye} J.,    {Yatawatta} S.,  2009, \mnras, 393, 32

\bibitem[\protect\citeauthoryear{{Tremblin}, {Audit}, {Minier} \&
  {Schneider}}{{Tremblin} et~al.}{2012}]{Tremblin2012}
{Tremblin} P.,  {Audit} E.,  {Minier} V.,    {Schneider} N.,  2012, \aap, 538,
  A31

\bibitem[\protect\citeauthoryear{{Truelove}, {Klein}, {McKee}, {Holliman} II,
  {Howell} \& {Greenough}}{{Truelove} et~al.}{1997}]{Truelove1997}
{Truelove} J.~K.,  {Klein} R.~I.,  {McKee} C.~F.,  {Holliman} II J.~H.,
  {Howell} L.~H.,    {Greenough} J.~A.,  1997, \apjl, 489, L179

\bibitem[\protect\citeauthoryear{{Van Loo}, {Butler} \& {Tan}}{{Van Loo}
  et~al.}{2013}]{VanLoo2013}
{Van Loo} S.,  {Butler} M.~J.,    {Tan} J.~C.,  2013, \apj, 764, 36

\bibitem[\protect\citeauthoryear{{Vazquez-Semadeni}}{{Vazquez-Semadeni}}{1994}]{Vazques1994}
{Vazquez-Semadeni} E.,  1994, \apj, 423, 681

\bibitem[\protect\citeauthoryear{{V{\'a}zquez-Semadeni}, {Gonz{\'a}lez},
  {Ballesteros-Paredes}, {Gazol} \& {Kim}}{{V{\'a}zquez-Semadeni}
  et~al.}{2008}]{Vazques2008}
{V{\'a}zquez-Semadeni} E.,  {Gonz{\'a}lez} R.~F.,  {Ballesteros-Paredes} J.,
  {Gazol} A.,    {Kim} J.,  2008, \mnras, 390, 769

\bibitem[\protect\citeauthoryear{{Wada} \& {Koda}}{{Wada} \&
  {Koda}}{2004}]{Wada2004}
{Wada} K.,  {Koda} J.,  2004, \mnras, 349, 270

\bibitem[\protect\citeauthoryear{{Wada} \& {Norman}}{{Wada} \&
  {Norman}}{2001}]{Wada2001}
{Wada} K.,  {Norman} C.~A.,  2001, \apj, 547, 172

\end{thebibliography}

\end{document}